\documentclass[aps,pre,twocolumn,superscriptaddress,longbibliography,floatfix]{revtex4-1}

\usepackage[utf8]{inputenc}
\usepackage[english]{babel}
\usepackage[T1]{fontenc}
\usepackage{lmodern}

\usepackage{amsmath,amsthm,amsfonts,amssymb,bm,amssymb}

\usepackage[colorlinks={true}, citecolor={blue}, filecolor={blue}, linkcolor={blue}, urlcolor={blue}]{hyperref}

\usepackage{graphicx}
\usepackage[caption=false]{subfig}
\graphicspath{{figs/}} 

\usepackage{microtype}
\usepackage{soul}

\newcommand{\Tr}{\operatorname{Tr}}

\newcommand{\kB}{k_\mathrm{B}}

\begin{document}

\title{Internal geometric friction in a Kitaev chain heat engine}

\author{Elif Yunt}
\email{elif.yunt@um.edu.mt}
\affiliation{Department of Physics, University of Malta, Msida MSD 2080, Malta}
\affiliation{Department of Physics, Ko\c{c} University, 34450 Sariyer, Istanbul Turkey}

\author{Mojde Fadaie}
\email{Seyedeh.mozhdeh.fadaie@umontreal.ca}
\affiliation{Department of Physics, Universite de Montreal, 1375 Avenue Thérèse-Lavoie-Roux, Montreal,Canada}
\affiliation{Department of Physics, Ko\c{c} University, 34450 Sariyer, Istanbul Turkey}

\author{\"{O}zg\"{u}r E. M\"{u}stecapl{\i}o\u{g}lu}
\email{omustecap@ku.edu.tr}
\affiliation{Department of Physics, Ko\c{c} University, 34450 Sariyer, Istanbul Turkey}

\author{Cristiane Morais Smith}
\email{c.demoraissmith@uu.nl}
\affiliation{Institute for Theoretical Physics, Utrecht University, Princetonplein 5, 3584 CC Utrecht, The Netherlands}


\begin{abstract}

	We investigate a heat engine with a finite-length Kitaev chain in an ideal Otto cycle.  It is found that the critical point of the topological phase transition coincides with the maxima of the efficiency and work output of the total Otto engine. Finite-size effects are taken into account using the method of Hill's nanothermodynamics, as well as using the method of temperature-dependent energy levels. We identify the bulk and boundary thermal cycles of the Kitaev chain engine and find that they are non-ideal Otto cycles. The physics of deviation from ideal Otto cycle is identified as a finite size effect, which we dub as  ``internal geometric friction'', leading to heat transfer from the bulk to the boundary during the adiabatic transformation of the whole system.  Besides, we determine the regimes allowing for independently running an ideal Otto refrigerator at the boundary and ideal Otto  engines in the bulk and in the whole system. Furthermore, we show that the first-order phase transition in the boundary and the second-order phase transition in the bulk can be identified through their respective contributions to the engine work output.

\end{abstract}

\maketitle
\section{Introduction}
 
Thermodynamic methods are strictly applicable only to describe energy processes of macroscopically large objects. A formulation of thermodynamics for finite-size systems has been proposed by T. L. Hill ~\cite{hillbook,doi:10.1002/ijch.196500008}, named as nanothermodynamics~\cite{e17010052,hill_perspective_2001,doi:10.1021/nl010027w}.  Thermodynamic behavior of bounded systems of current interest, in particular topological insulators and topological phase transitions (TPTs), can be examined by the nanothermodynamic approach ~\cite{quelle,universality,broeke_thermodynamic_2018}. A topological phase transition separates zero temperature quantum states of a system distinguished by topological numbers, which are invariant for the system with an energy gap under any adiabatic changes. Gap closing is a necessary but not sufficient condition for the TPT, where the topological number becomes ill-defined. When the bulk has non-zero topological number, gapless edge states emerge at the boundary of the system in accordance with the bulk-boundary correspondence. The bulk- boundary correspondence leads the boundary of the topological system to manifest edge states due to the non-trivial topology of its bulk.  Topological numbers for pure, zero-temperature, states are determined by the so-called Berry phase. Extension of topological phases to finite temperatures can be done by the generalizations of the Berry phase, for example, the so-called Uhlmann phase~\cite{delgado1,delgado2}. Intriguing thermal behavior, such as chiral heat currents that flow against the thermal gradient~\cite{RivasDelgado2017}, and signatures of the TPTs at the work output of quantum heat engines~\cite{ PhysRevE.98.052124} have been reported in topological systems.

Here, we ask whether the gapless edge modes and the gap-closing bulk states at the TPTs contribute to the work output of a nanothermodynamic heat engine with distinct signatures. We explore this question specifically for a finite-length Kitaev chain~\cite{ Kitaev_2001}, which hosts topological zero-energy Majorana modes localized at the edges. The Majorana zero modes are appealing for topological quantum computation~\cite{KITAEV20032}. Due to the challenges concerning their direct observation, indirect probing schemes of these modes have been discussed in~\cite{Alicea_2012}. An interesting proposal is to utilize stroboscopic heat current in a time-periodic modulated spin chain to detect signatures of Floquet-Majorana modes ~\cite{PhysRevB.96.125144}. It is known that the Kitaev chain undergoes a second-order phase transition at the bulk and a first-order phase transition at the boundary~\cite{quelle}. We can identify the order of the topological phase transition in the work output of the topological heat engine by drawing an analogy of the boundary and bulk contributions with the quantum phase transitions of the Landau-Zener model in the cases of the level crossing and avoided crossing, respectively.

Besides, we explore the energy exchange between the bulk and the edge. Since the early studies performed by Leonardo da Vinci, it is traditionally accepted that the friction force between two identical objects is independent of their geometry – i.e. shape and size. Modern tribology studies, however, have revealed that the bulk of the sliding bodies, in addition to the interface, plays also a role on the friction~ \cite{PhysRevX.6.041023,Ghosh_2017}. The effects of orientation, shape, and size on friction can be significant at any length scale, ranging from earthquake fractures to nanoparticles~\cite{Vanossi_2018} or biological cells~\cite{PhysRevLett.111.198101}. By introducing holes in the systems, one can show that friction is sensitive to the topological changes~\cite{Dodou}. Our study reveals an “internal” type of geometrical friction, that emerges between the bulk and the boundary of a finite-size object. Due to its size dependence, we dub it  “internal geometric friction”. It exhibits sensitivity to the topological character of the system, including the order of a topological phase transition during the “sliding”. Here, the “sliding” refers to the different evolutions of bulk and boundary subsystems when an adiabatic transformation is applied to the state of the whole system. It can be envisioned as a “relative motion” of the bulk and the boundary on a parameter trajectory. Along the parameter path, energy and entropy exchange occurs between the bulk and the boundary, so that dissipation due to internal geometric friction reduces the efficiency of the work done on the bulk of the object along the parametric path. A perfect transformation on the bulk would require an infinite size and an infinite time. The term internal friction is used to describe imperfect finite-time adiabatic transformations, which builds up coherence in the internal degrees of freedom of a system. They can be eliminated using the so-called shortcuts to the adiabaticity. We leave the questions of bulk-boundary coherence, explicit isolation of topological contribution, and shortcuts in internal geometric friction questions open for future investigations.

Furthermore, we address the influence of a possible interface that connects the heat baths to the Kitaev chain on the heat transfer. The latter question is investigated by another finite-size thermodynamics method, known as temperature-dependent energy levels (TDELs) ~\cite{Elcock_1957,Rushbrooke,miguel1,miguel2,miguel_temperature-dependent_2015} [17-21]. The TDELs scheme takes into account the contribution of the bath-system interface as another energy dissipation channel~\cite{yamano2,yamano_effect_2017}, consistent with the extended second law of thermodynamics~\cite{Shental_2009}.

Existing atomic and spin quantum heat engines are mainly of academic interest, producing too small work output for any practical application. Here, we consider a many-body engine with potential to produce much more significant work output ($\sim$meV) in comparison to the single spin heat engine ($\sim$peV). Coupling the topological system to a mechanical, optical, or a magnetic “flywheel”, can also be envisioned for practical device application. Further enhancement, as we show here, is possible when the  topological phase changes during the engine cycle. Our results on the role of interface in imperfect heat transfer to the bulk due to boundary, or internal geometric friction, could be significant on such future applications, as well as to interpret existing experiments~\cite{broeke_thermodynamic_2018}. Elaboration of these practical questions are beyond the scope of our present contribution.

This paper is organized as follows.  We briefly review the Hill's nanothermodynamics and the framework of TDELs that we use in two subsections in Sec.~\ref{sec:methods}. Our model system, the finite-length Kitaev Chain is presented in Sec.~\ref{sec:model}. The results and discussions are given in Sec.~\ref{sec:res} in four subsections. We conclude in Sec.~\ref{sec:conc}. In the  Appendix, we elaborate on the regimes allowing for independently running an ideal Otto refrigerator at the boundary and ideal Otto  engines in the bulk and in the whole system.

\section{Methods}\label{sec:methods}

We summarize here the key points of the two major historical approaches developed to extend the thermodynamic description to finite size systems: (i) Hill's nanothermodynamics~\cite{hillbook} and (ii)TDELs~\cite{miguel1,yamano_effect_2017,yamano2}. We shall restrict ourselves to these two general methods to develop our results. There are also other approaches to describe the thermodynamics of small systems, but they are beyond the scope of the present paper~\cite{lebowitz_thermodynamic_1961,sisman,bedeaux_hills_2018}. 
\subsection{Hill's Nanothermodynamics}\label{sec:nanothermo}
\begin{figure}[htb!]
	\centering
	\includegraphics[width=8.3 cm]{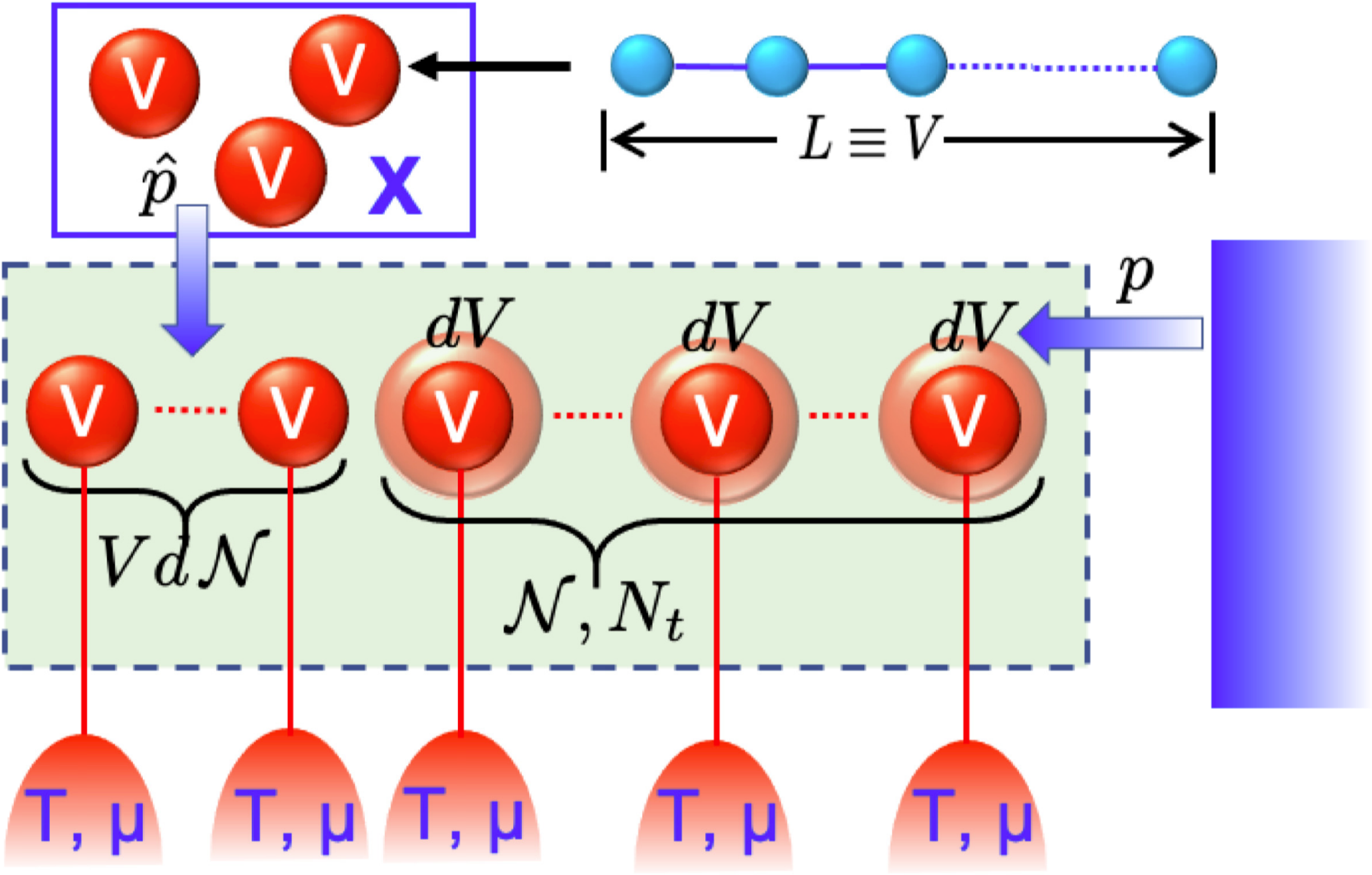}
	\caption{(Color Online) Schematic description of the idea of the Hill's thermodynamics. Identical, non-interacting ${\cal N}$ replicas of a system of finite volume $V$ form a macroscopic ensemble. Each copy is at temperature $T$. The total number of particles is $N_t$, which can change at the cost of chemical potential $\mu$, same for each replica. At the cost of energy $X$, associated with so-called integral pressure $\hat p$, another system can be added to the ensemble. Addition of $d{\cal N}$ systems at constant volume $V$ increases the ensemble volume by $Vd{\cal N}$. The total volume can also be varied by an external work, associated with mean pressure $p$, changing the size of each system by an amount of $dV$ at constant ${\cal N}$. We will apply this framework on a finite-size system, a Kitaev chain of length $L$.}
	\label{fig:HillSketch}
\end{figure}

Laws of thermodynamics are stated for macroscopic systems. In this section, we describe the so-called Hill's nanothermodynamics, which was developed to extend thermodynamical laws to the regime of finite-size
systems~\cite{hillbook,doi:10.1002/ijch.196500008}. The idea starts by envisioning a thermodynamically large ensemble of small objects of volume $V$, as sketched in Fig.~\ref{fig:HillSketch}. The total system comprises identical copies of the finite-size system of interest, which are not interacting with each other. We can apply ordinary thermodynamics on the large ensemble to express its internal energy.

Let us assume that each replica is at temperature $T$. Heat contribution to the internal energy is given by
$TdS_t$, where $S_t$ is the entropy of the whole ensemble. The number of particles in each copy can change at the energy cost of $\mu$. Denoting the total number of particles by $N_t$ in the ensemble,
the chemical work contributes to the internal energy by an amount of $\mu dN_t$. While these two terms are quite standard, the volume changes become
non-trivial. When the number of systems ${\cal N}$ in the ensemble fluctuate by $d{\cal N}$ at constant $V$, the total volume varies by $Vd{\cal N}$. Introducing the so-called integral pressure $\hat p$ on the system, the internal energy change by such a process can be written as $-\hat pVd{\cal N}$.
The total volume could also be changed at a constant ${\cal N}$  by applying pressure $p$ on replicas to change their sizes by $dV$.
This would contribute an energy of $-p{\cal N}dV$. Summing up, the Gibbs differential relation for the internal energy of the whole ensemble reads
\begin{equation}\label{eq:dEt}
dE_t=TdS_t+\mu dN_t-p{\cal N}dV+X d{\cal N},
\end{equation}
where $X=-\hat pV$, can be interpreted as the energy cost of adding another small-system replica to the ensemble.

Let us impose grand canonical ensemble, constant $\mu,V,T$, conditions on the system and each of its copies so that Eq.~(\ref{eq:dEt}) simplifies to 
\begin{equation}\label{eq:dEt2}
dE_t=TdS_t+\mu dN_t+X d{\cal N}.
\end{equation}
Integrating Eq.~(\ref{eq:dEt2}) by the Euler theorem, for fixed $T,\hat p,V,\mu$ we find 
\begin{equation}\label{eq:hill-E}
E_t=TS_t+\mu N_t+X{\cal N}.
\end{equation}
A critical step is to take ensemble averages per system, 
$E:=E_t/{\cal N}$ and $N:=N_t/{\cal N}$, as the mean internal energy and mean number of particles for a single system, respectively. We remark that the entropy $S:=S_t/{\cal N}$ is the same for each small system~\cite{hillbook,doi:10.1002/ijch.196500008}.
Introducing $\epsilon:=(p-\hat{p})V$, so-called subdivision potential, we then write
\begin{equation}\label{eq:hill-Esys}
E=TS+\mu N-pV+\epsilon.
\end{equation}
Remarkably, this internal energy expression cannot be found by Euler integration of the corresponding Gibbs differential relation for the internal energy of a single system. Substituting $E_t={\cal N}E, N_t={\cal N}N, S_t={\cal N}S$ in Eq.~\ref{eq:dEt}, we get
\begin{equation}\label{eq:hill-dEsys}
dE=TdS+\mu dN-pdV,
\end{equation}
without any $\epsilon$. The underlying result is that the internal energy for a finite-size system is no longer a linear homogeneous function of $S,V$, and $N$ so that Euler integration cannot be applied. Physically, this corresponds to the breaking of the extensivity property of thermodynamic functions at a small system size. The ordinary thermodynamics can be recovered in the limit $\epsilon\rightarrow 0$, which is the case for large systems.

Using Eqs.~(\ref{eq:hill-Esys}) and (\ref{eq:hill-dEsys}),
the generalized Gibbs-Duhem equation can be written as
\begin{equation}\label{eq:hill-gibbsDuhem}
d\epsilon=-SdT+Vdp-Nd\mu.
\end{equation}
With the definition of $\epsilon$, this yields the differential relation for $X$,
\begin{equation}\label{eq:hill-deps}
dX=-SdT-pdV-Nd\mu,
\end{equation} 
which is the same as the ordinary thermodynamic one,
by identification of $X=-\hat p V\equiv \Phi$ as 
the grand potential of a single system. Using the grand potential allows us
to establish the connection of Hill's framework to the statistical mechanics through
\begin{equation}\label{eq:statMechConnect}
\Phi(T,\mu,V) = - k_BT\ln \Xi(T,\mu,V),
\end{equation}
where 
\begin{equation}
\Xi(\mu,T,V)=\Tr\exp[-\beta(H-\mu N)],
\end{equation}
is the grand canonical partition function 
for a single system described by the Hamiltonian $H$ in the ensemble
with $\beta=1/k_BT$. Note that we can split the
grand canonical partition function into two parts,
$\Phi=-pV+\epsilon$, demonstrating the subdivision potential as the difference between the grand potentials of the bounded and unbounded systems. Accordingly, the possibility to describe boundary effects on the thermodynamic behavior of small systems emerges naturally in Hill's framework. On the other hand, in contrast to $\Phi$, a closed-form statistical mechanical expression or microscopic origin for the subdivision potential $\epsilon$ is not given in the Hill's nanothermodynamics.

Equations (\ref{eq:hill-Esys})-(\ref{eq:hill-gibbsDuhem}) constitute the essential relations of Hill's nanothermodynamics. We remark two subtle points: first, it is assumed that the small systems can be thermalized to the bath temperature $T$, and Hill's thermodynamics does not give any mechanism for this thermal equilibration; second, the subdivision potential is a phenomenological term presented without any microscopic origin. These two issues are going to be addressed within the framework of the TDELs method. A special method to construct $\epsilon$ using statistical mechanical calculation of $\Phi$ has been introduced in Ref.~\cite{quelle,universality}, and will be described subsequently.
\subsection{Temperature-dependent energy levels}\label{sec:tdel}
The method of TDELs has been proposed as a fast and convenient way to perform statistical mechanic calculations for an assembly of systems~\cite{Rushbrooke,Elcock_1957}. It has been
applied to semiconductors
~\cite{VARSHNI1967,PhysRevB.30.5766,Allen_1976,Patrick_2014}, superfluids~\cite{PhysRevLett.119.256802}, optomechanical oscillators~\cite{kolar_optomechanical_2017,kolar2}, heat losses in thermoelectric systems~\cite{yamano2,yamano_effect_2017}, and thermalization
of finite-size systems~\cite{miguel1,miguel2,miguel_temperature-dependent_2015}. 
 Even before TDELs were proposed,
advantages of temperature dependent mean
potentials and corresponding forces were known in statistical mechanics of fluid mixtures~\cite{kirkwood_statistical_1935}.
A modern application of TDELs connects them to a temperature dependent effective Hamiltonian, so-called ``Hamiltonian of mean force'' to explore strong coupling
of a small system to a heat bath ~\cite{seifert_first_2016,talkner_open_2016}, where counterintuitive negative thermophoresis effects can be explained using TDELs~\cite{de_miguel_negative_2019}. 

\begin{figure}[t!]
	\centering
	\includegraphics[width=8.3 cm]{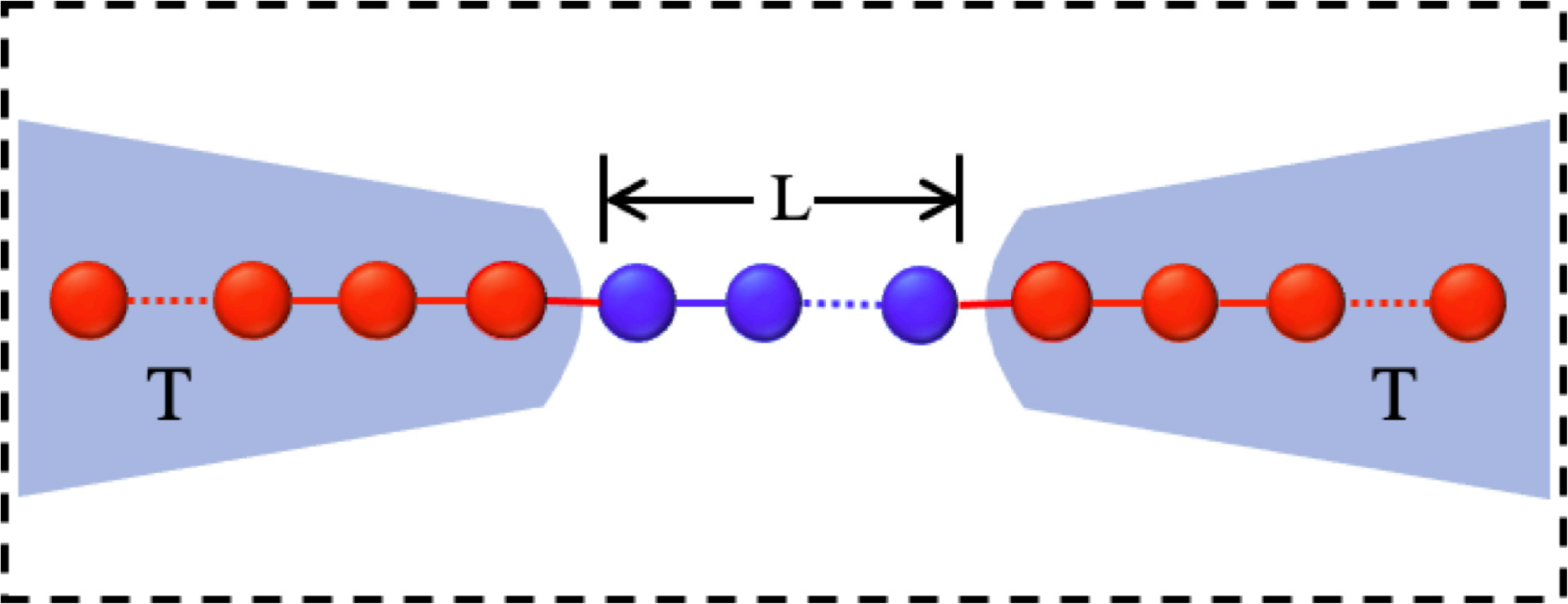}
	\caption{(Color Online) Illustration of the method of temperature dependent energy levels (TDELs) applied to a finite-length (L) chain. Total system of the chain and the semi-infinite interfaces is macroscopic and hence can be described by ordinary thermodynamics. The total system is assumed to be in a thermal equilibrium state due to the exposure of the leads to a thermal environment at temperature $T$. If we are interested only in the central chain, then interface degrees of freedom can be integrated out from the state of the total system. Such prior averaging by partial integration yields a temperature dependent reduced state for the finite chain. The TDELs are conveniently introduced by expressing such a state in a canonical thermal (Gibbs) form. The TDELs approach differs from that of the Hill (cf. Fig.~\ref{fig:HillSketch}) in its construction of the macroscopic system in terms of an interface. In the Hill's nanothermodynamics, an ideal "gas" of replicas of the finite system is used to be able to apply ordinary macroscopic thermodynamics.}
	\label{fig:TDELsketch}
\end{figure}

Temperature dependent effective Hamiltonians and  TDELs arise after a prior averaging over certain possible
microstates of the assembly in thermal equilibrium , as illustrated in Fig.~\ref{fig:TDELsketch}.
 The original definition of TDELs~\cite{Elcock_1957} and the definition of Hamiltonian of mean force~\cite{seifert_first_2016,talkner_open_2016}
can be combined by the expression 
\begin{eqnarray}\label{eq:TDELdefn}
\langle n|\frac{\text{e}^{-\beta H_a(T)}}{Z_a}|n\rangle:=&&\frac{\text{e}^{-\beta E_n(T)}}{Z_a} \nonumber  \\
  \equiv && \langle n| \text{Tr}_b 
\left(\frac{\text{e}^{-\beta H}}{Z}\right)|n\rangle.
\end{eqnarray}
Here $\beta=1/k_BT$ with $k_B$ being the Boltzmann constant. $H=H_a+H_b+V_{ab}$ is the Hamiltonian of a macroscopic system in thermal equilibrium. For clarity and brevity of notation we restricted it to an assembly of two parts labeled with $a$ and $b$.
Part-$a$ is the system we want to single out so that partial integration, or trace, is employed over the quantum numbers associated with part-$b$. In Fig.~\ref{fig:TDELsketch} for example, 
the central small chain corresponds to part-$a$, while the macroscopic thermal leads would be part-$b$. 
Quantum numbers associated with the part-$a$ are represented by the symbol $n$. We remark that different definitions of the Hamiltonian of the mean force can be found in literature, due to dropping or partially keeping the partition functions~\cite{seifert_first_2016,talkner_open_2016,de_miguel_negative_2019}, 
\begin{eqnarray}\label{eq:Za}
Z_a&:=&\text{Tr}_a \text{e}^{-\beta H_a(T)}= \sum_n\text{e}^{-\beta E_n(T)},\\
Z&:=&\text{Tr}_b \text{e}^{-\beta H},
\end{eqnarray}
of the part-$a$ and the total system, respectively. They contribute to the Hamiltonian of the mean force as constant potential shifts.

The definition of the TDEL as well as of the Hamiltonian of the mean force is not unique. The choice of monotonically decreasing exponential, or the canonical Gibbs form, for the reduced density matrix of part $a$ is to allow to recover conveniently the standard thermodynamic 
expression for the free energy~\cite{Rushbrooke,Elcock_1957},
\begin{eqnarray}\label{eq:Fs}
F_a=-k_BT\ln{Z_a}=-k_BT\ln{\sum_n\text{e}^{-\beta E_n(T)}}.
\end{eqnarray}
On the other hand, thermodynamically consistent calculations of the entropy $S_a$ and the mean energy $E_a$
\begin{eqnarray}\label{eq:Fs}
E_a&=&k_BT^2\frac{\partial \ln Z_a}{\partial T},\\
S_a&=&\frac{U_a}{T}+k_B\ln{Z_a},
\end{eqnarray}
lead to the expressions
\begin{eqnarray}
U_a&=&\sum_np_a(n)\Big[E_n(T)-T\frac{\partial E_n(T)}{\partial T}\Big],\label{eq:Ua}\\
S_a&=&-k_B\sum_np_a(n)\Big[\ln{p_a(n)}+\frac{\partial E_n(T)}{k_B\partial T}\Big],\label{eq:Sa}
\end{eqnarray}
where $p_a(n):=\exp{(-\beta E_n(T))}/Z_a$.

These expressions indicate that in the presence of TDELs, the energy transferred from the heat bath to the system cannot be completely identified as heat. To see this clearly, let us rewrite Eq.~(\ref{eq:Ua}) as
\begin{eqnarray}
U=\langle H\rangle-T\left\langle
\frac{\partial H}{\partial T}\right\rangle,
\end{eqnarray}
where $\langle H\rangle$ refers to the usual thermodynamic expression of the mean energy, given by the first term in Eq.~(\ref{eq:Ua}). We dropped the label $a$ for brevity. 
In differential form $U$ becomes,
\begin{eqnarray}
dU=\sum_n (E_ndp_n+p_ndE_n)
-d\left\langle\frac{\partial H}{\partial T}\right\rangle.
\end{eqnarray}
Similarly, differential form of Eq.~(\ref{eq:Sa}) gives
\begin{eqnarray}
TdS=\sum_n E_ndp_n
-Td\left\langle \frac{\partial H}{\partial T}\right\rangle.
\end{eqnarray}
Combination of the last two expressions yields the first law of
thermodynamics in the presence of TDEL,
\begin{eqnarray}
dU = dW+TdS
-\left\langle \frac{\partial H}{\partial T}\right\rangle dT,
\end{eqnarray}
with
\begin{eqnarray}
dW=
\sum p_n dE_n.
\end{eqnarray}
In the case of heating, where a heat bath or an interface lead
is attached to the small system, we take $dW=0$ and write~\cite{yamano2,yamano_effect_2017,miguel1,miguel_temperature-dependent_2015}
\begin{equation}\label{eq:dQeff}
\delta Q_{\rm eff}=\delta Q-\left\langle \frac{\partial H}{\partial T}\right\rangle dT.
\end{equation} 
This expression tells us that some of the heat, 
$\delta Q$  from the bath
has been lost at the interface in terms of work, described
by the second term in Eq.~(\ref{eq:dQeff}), and the system received
a reduced amount of heat $\delta Q_{\rm eff}$.
As the system temperature changes to match with that of the heat bath, the energy gaps between the energy levels may change, in addition to their  populations. Accordingly, the second term cannot be interpreted as heat; it represents the work applied to change the energy gaps. Such a mechanism has been discussed in detail to explain thermalization of small systems~\cite{miguel1,miguel2,miguel_temperature-dependent_2015}

Eq.~(\ref{eq:dQeff}) originates from the first law modified for a small system; rewriting it as $dS_T=dQ/T-\langle \partial H/\partial T\rangle dT/T$, one recovers the modified second law of information transfer channels~\cite{Shental_2009}. The entropy transferred from a heat source into the system through the boundary is balanced by the entropy lost at the boundary. The boundary acts as an energy channel interfacing the  heat source and the system. 

Hill's nanothermodynamics and the method of TDELs can be connected to each other by recognizing that the subdivision potential can be determined by~\cite{miguel1}
\begin{equation}\label{eq:Hill-TDEL-link}
\epsilon=T\left\langle \frac{\partial H}{\partial T}\right\rangle.
\end{equation}
This can be seen by substituting $p(n)=\exp{[-\beta (E_n(T)-\mu)]}/\Xi$ into the first term in Eq.~(\ref{eq:Sa}), which yields
\begin{eqnarray}\label{eq:HillTDELclarify}
\langle H\rangle=TS+\mu N-k_BT\ln \Xi+T\left\langle\frac{\partial H}{\partial T}\right\rangle.
\end{eqnarray}
Hill's approach determines the mean energy from the usual thermodynamic expression so that $E\equiv\langle H\rangle$ in Eq.~(\ref{eq:hill-Esys}); while $S$ is identified with
that of TDEL~\cite{miguel1}. Recognizing $k_BT\ln \Xi=\hat{p} V$ and using
$\epsilon=(p-\hat{p})V$ in Eq.~(\ref{eq:HillTDELclarify}), Eq.~(\ref{eq:Hill-TDEL-link}) follows.

We can use the calculations of Hill's nanothermodynamics to evaluate the heat transfer as described by
the TDEL method according to
\begin{equation}\label{eq:Qeff}
Q_{\rm eff}=\int T dS - \int\frac{\epsilon}{T}dT.
\end{equation} 
We emphasize that the first term is written in terms of the $S$ of Hill's nanothermodynamics, according to which we have $dE=dQ=TdS$. The TDELs method yields a correction term associated with the work done on
the energy gaps, such that some heat is dissipated at the system-bath interface and only an effective heat is received by the system.

Next, we apply this generic formalism to a specific model. 
 \section{Model: Kitaev Chain in Otto Cycle}\label{sec:model}
The Kitaev chain is a one-dimensional topological superconductor model~\cite{Kitaev_2001}, which consists of spinless fermions, described by the Hamiltonian
\begin{eqnarray}\label{eq:HKC}
	H=-\mu\sum_{i=1}^n a^\dagger_{i}a_{i} 
	-\sum_{i=1}^{n-1} \Big(t a^\dagger_{i} a_{i+1}-
	\Delta a_{i+1} a_{i} + h.c.\Big).
\end{eqnarray}
Here, $i=1...n$ labels the $n$ sites on the one-dimensional finite-length Kitaev chain, $\mu$ is the chemical potential, $t$ is the hopping parameter, and $\Delta$ is the superconducting pairing parameter. The fermionic annihilation (creation) operators, $a_{i}(a^\dagger_{i})$ satisfy the anti-commutation relations $\{a_{i},a^\dagger_{j}\}=\delta_{ij}$.
The Kitaev chain undergoes a topological phase transition at $|\mu|=2t$, which separates a topological phase for 
$|\mu|<2t$ from a trivial phase for $|\mu|>2t$. In the topological phase, the Kitaev chain hosts a pair of Majorana zero modes at its edges. 

Following the Hill's nanothermodynamics, an ensemble of 
identical, equivalent, and non-interacting copies of a finite-length
Kitaev chain will be considered. For the TDEL approach,
we consider a macroscopic Kitaev chain decomposed into two parts: One part is taken as the finite-length system, and the other is an interface lead, which is long enough to be assumed in thermal equilibrium with a heat bath~\cite{Elcock_1957}. TDELs arise as a result of averaging over the interface microstates, and describe the heat exchange during the thermalization of the finite-length chain with the heat bath in a thermodynamically consistent way. 

Thermodynamical properties of bulk and boundary of the finite-length Kitaev chain can be investigated using Hill's nanothermodynamic framework from the statistical mechanical point of view. The  idea described in Ref.~\cite{quelle,universality} is to use a  linear fit to the grand potential in Eq.~(\ref{eq:statMechConnect}). The chemical potential term is already included in the Kitaev model, and hence the grand canonical partition function reduces to
\begin{equation}
	\Xi(\mu,T,V)=\Tr\exp[-\beta H].
\end{equation}
The chemical potential term in the Kitaev model makes two contributions to the grand potential. It gives a constant shift, which can be dropped from the grand potential, but is also present in the eigenvalues of the Kitaev model in Eq.~(\ref{eq:HKC}). Accordingly, the grand potential is dependent on the chemical potential.
The following ansatz describes the linear fit for the grand potential~\cite{quelle}
\begin{equation}
\Phi(\mu,T,L)=\Phi_c(\mu,T)L+\Phi_0(\mu,T).
\label{ansatz1}
\end{equation}
Here, L is the total length of the chain, and $\Phi_cL=-pL$ is the bulk grand potential, which is extensive, while
$\Phi_0=\epsilon$ is the subdivision potential, which emerges due to the finite length of the chain. The corresponding entropies obey a similar relation,
\begin{equation}
S = S_c L+ S_0.
\label{eq:Sbulkedge}
\end{equation}

The first term, the bulk entropy, coincides with the entropy of the total system in the thermodynamic limit, where $S_0$ disappears at $L\rightarrow\infty$. This first term can be larger than the entropy of the total finite system because $S_0$ can be negative. A negative $S_0$ is allowed in the Hill's thermodynamics, which is a non-extensive thermodynamic theory. Similar conclusions apply to other thermodynamic quantities, which are extensive in the thermodynamic limit but non-extensive in the finite-size nanothermodynamic regime.

We first find the eigenvalues of the Hamiltonian in Eq.~(\ref{eq:HKC}) through a numerical calculation, then evaluate the total entropy
$S$ of the chain. $S_c$ and $S_0$ are determined by using a linear fit to $S$ for an $n$-site chain (we take the unit length of the chain as $1$ so that $L=n$) in the interval $200<n<225$. The length is chosen to be sufficiently
large to make the linear fit a valid approximation~\cite{quelle}. 
Repeating the procedure for different temperatures, the $T-S$ relation is found.
We will consider the thermodynamic cycle of an ideal Otto engine. The Otto cycle consists of two isentropic (adiabatic) and two isochoric (isoparametric) stages \cite{Quan2007}.
We take the chemical potential $\mu$ as the control parameter for
the Otto cycle. The curves in the $T-S$ plane at different $\mu$ will be used to determine the explicit cycle diagram 
and the associated work output. In the following subsections, the corresponding work and efficiency of the Kitaev chain Otto cycle will be investigated by distinguishing the 
bulk and boundary contributions and by considering the effect of bath-system interface as an energy channel, which 
modifies the heat exchange between the heat baths and the Kitaev chain.

A Kitaev chain could be realized in a trapped ion quantum simulator~\cite{Mezzacapo_2013}. Such a system has the advantage of single site resolution for heterogenous probing of thermal properties.  Accordingly, either the bulk or the boundary of the ion string could be thermally excited. For that aim, one could use a photonic quantum interface to encode thermal Gibbs states locally. Different work and heat transfer at the bulk and boundary could be investigated by spatially resolved measurements. Other quantum simulators for topological matter, such as superconducting chains~\cite{Hu2017} or optical lattices~\cite{PhysRevA.98.033604,PhysRevLett.91.090402}, or arrays of nonlinear optical cavities~\cite{PhysRevLett.109.253606} are also possible embodiments where bulk and boundary effects could be experimentally distinguished. One can envision local measurements on bulk and boundary by using tunnel-gates~\cite{Zhang2018}, narrow local gates~\cite{Das2012, Mourik1003} or high-resolution scanning-tunneling microscopy/spectroscopy~\cite{Wang333}, which are utilized for electrical transport experiments to probe Majorana zero-modes. While such possible experiments would be of fundamental interest, they can eventually lead to practical quantum thermal-topological devices, too.
A robust topological heat engine, which exploits  boundary properties of the one dimensional nonequilibrium Su-Schriefer-Hegel model has been proposed in a thermoelectric set-up in Ref.~\cite{Bohling_PhysRevB.98.035132}.  These thermo-topological devices may enable probing intriguing effects like a Berry-phase  induced heat pumping~\cite{Ren_PhysRevLett.104.170601} or a geometric phase induced heat flux~\cite{Wang_PhysRevA.95.023610}.	
Thermal devices such as heat pumps using braiding of Majorana zero modes have also been recently proposed~\cite{PhysRevB.99.205101}.
 
\section{Results and Discussion}\label{sec:res}
In our calculations, we consider a Kitaev chain of length $n=225$ with a superconducting pairing  $\Delta=0.25$ and a hopping  $t=0.25$. $\mu$ is used as the control parameter, for which a TPT takes place at $\mu=0.5$. We use the same set of parameters as in Ref. ~\cite{universality}.
\subsection{Work and efficiency of finite-length Kitaev chain Otto engine} \label{sec:Total}
\begin{figure}[t!]
	\centering
	\includegraphics[width=8.3 cm,height=5.9cm]{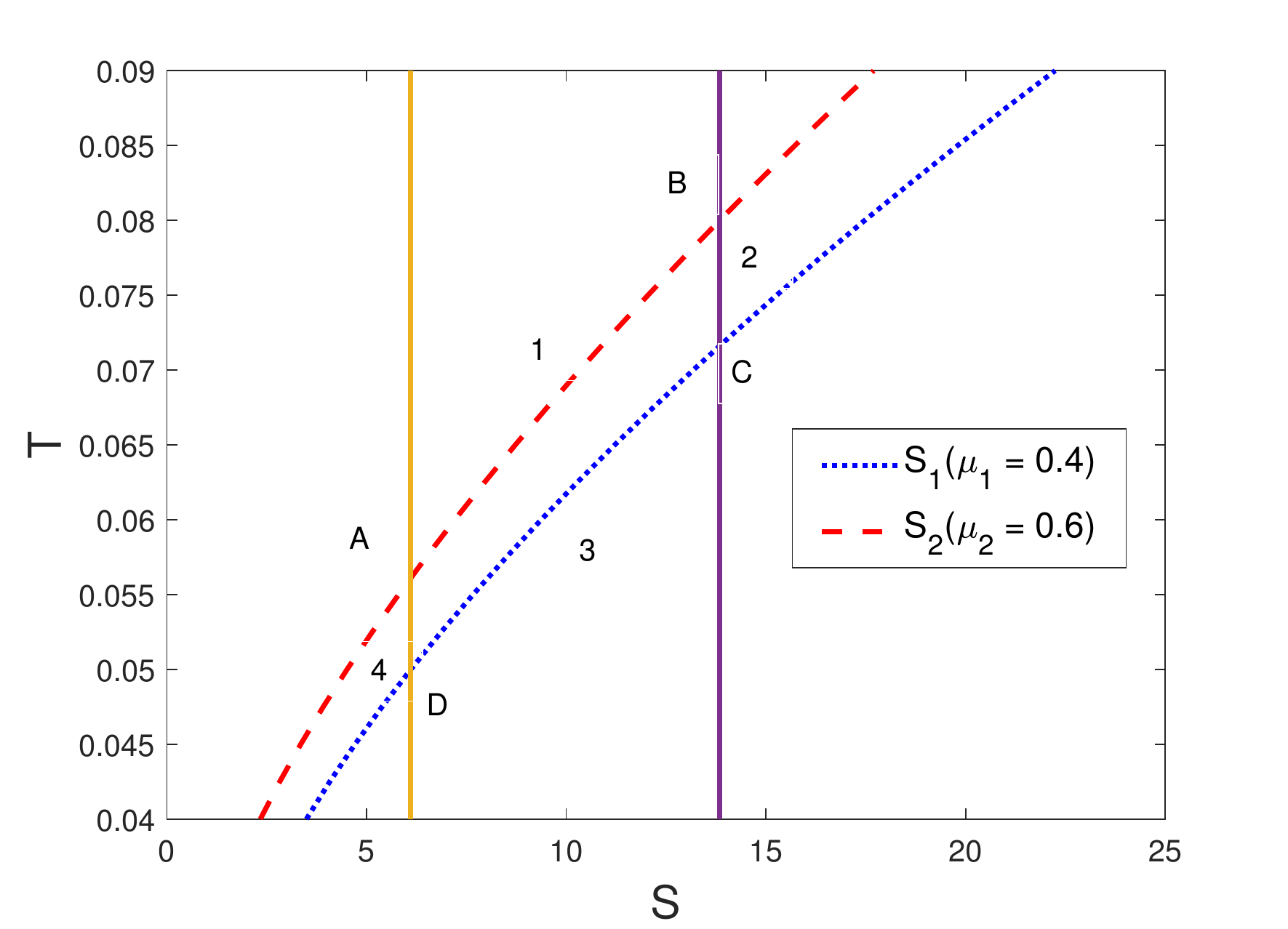}
	\caption{(Color Online) Temperature-Entropy ($T-S$) diagramme of a finite-length Kitaev chain according to Hill's nanothermodynamics. The chain has $n=225$ sites and is characterized by a superconducting pairing parameter $\Delta=0.25$ and hopping parameter $t=0.25$. The blue dotted curve is for the chemical potential $\mu_1=0.4,$ while the red dashed one is for $\mu_2=0.6$. An  Otto cycle can be defined using the segments between the points $A,B,C,$ and $D$  determined by the intersection of the curves and the solid yellow and solid purple constant entropy  lines. The cycle operates between a hot bath at temperature $T_B = 0.08$ and a cold bath at temperature 
	$T_D = 0.05$ in the clockwise direction. The stages of the cycle are numbered $1, 2, 3, 4$ in order.}
	\label{fig:ST1}
	\end{figure}

Let us first consider the finite-length Kitaev chain as a whole.
The curves in the $T-S$ plane of the chain at $\mu_1=0.4$ (blue dotted) and $\mu_2=0.6$ (red dashed)  are plotted in Fig.~\ref{fig:ST1}. 
The vertical lines are the constant entropies $S_2(T_A)=S_1(T_D)$
and $S_2(T_B)=S_1(T_C)$, where $T_x$ denote the temperature of the chain at the point $x=A,B,C,D$. The lines are fixed by the hot and cold bath temperatures, $T_B = 0.08$  and $T_D = 0.05$, respectively. 

The Otto cycle is then determined by finding the 
intermediate temperatures, $T_A$ and $T_C$ from the constant entropy conditions. By intermediate temperatures, we refer to the temperatures the total system attains between the hot and cold baths.  We note that we work in the low-temperature regime, where the temperatures in the Otto cycle are much less than the energy gap.  The system goes through the following stages in the Otto Cycle:
\begin{itemize}
	\item Stage 1 ($A \rightarrow B$): The system is initially at temperature $T_A,$ and is coupled to a hot bath. It thermalizes to the hot bath temperature $T_B=0.08$ through an isoparametric process, where the chemical potential is $\mu = \mu_2 = 0.6.$ During this process, no work is done. The amount of heat injected to the system is given by 
	\begin{equation}\label{eq:heatAB}
		Q_{\text{AB}}= \int_{T_A}^{T_B}T\frac{dS_2}{dT}dT.
	\end{equation}
	\item Stage 2 ($B \rightarrow C$):  This is an adiabatic process. The entropy of the system remains constant. The system is separated from the hot bath. As the chemical potential changes from $\mu=\mu_2$ to $\mu=\mu_1,$ the chain attains the intermediate temperature $T_C$ at point C.  There is no heat exchange in this process but work is done on the system. The constant entropy condition is $S_2(T_B)=S_1(T_C).$
	\item Stage 3 ($C \rightarrow D$): This is an isoparametric stage, where $\mu_1$ is kept constant. The system is brought into contact with a cold bath at temperature $T_D$. As the chain thermalizes to the cold bath temperature, heat is ejected from the system. The ejected heat is given by
	\begin{equation}\label{eq:heatCD}
	Q_{\text{CD}}= \int_{T_C}^{T_D}T\frac{dS_1}{dT}dT.
	\end{equation}
	\item Stage 4 ($D \rightarrow A$): The system goes through another adiabatic process from point D to point A, in which the constant entropy condition is  $S_1(T_D)=S_2(T_A).$ The chemical potential is changed from $\mu_1$ to $\mu_2.$ There is no heat exchange and work is done by the system.
	
\end{itemize}

 We use a sign convention, where heat injected to (ejected from) the system is always taken to be positive (negative). We note that the sign convention for heat and work are different such that positive work signifies that the system does work on an  external agent, whereas positive heat means heat is injected into the system as aforementioned. The heat exchanges of the chain with the heat baths at the isoparametric stages $A\rightarrow B$ and $C\rightarrow D$  are calculated by Eqs. (\ref{eq:heatAB}-\ref{eq:heatCD}),
where the entropies of the chain for $\mu=\mu_1$ and $\mu=\mu_2$ are distinguished by $S_1$ and $S_2$, respectively.
The net work performed by the cycle is then calculated by
$W=Q_{\text{AB}}+Q_{\text{CD}}$. 

 $Q_{\text{AB}}>0>Q_{\text{CD}}$ and the positivity of $W$ is required
for heat engine operation, while $Q_{\text{CD}}>0>Q_{\text{AB}}$ and a negative work output would be
the case of refrigerator behavior.

\begin{figure}
	\centering
	\includegraphics[width=8.3 cm,height=5.9cm]{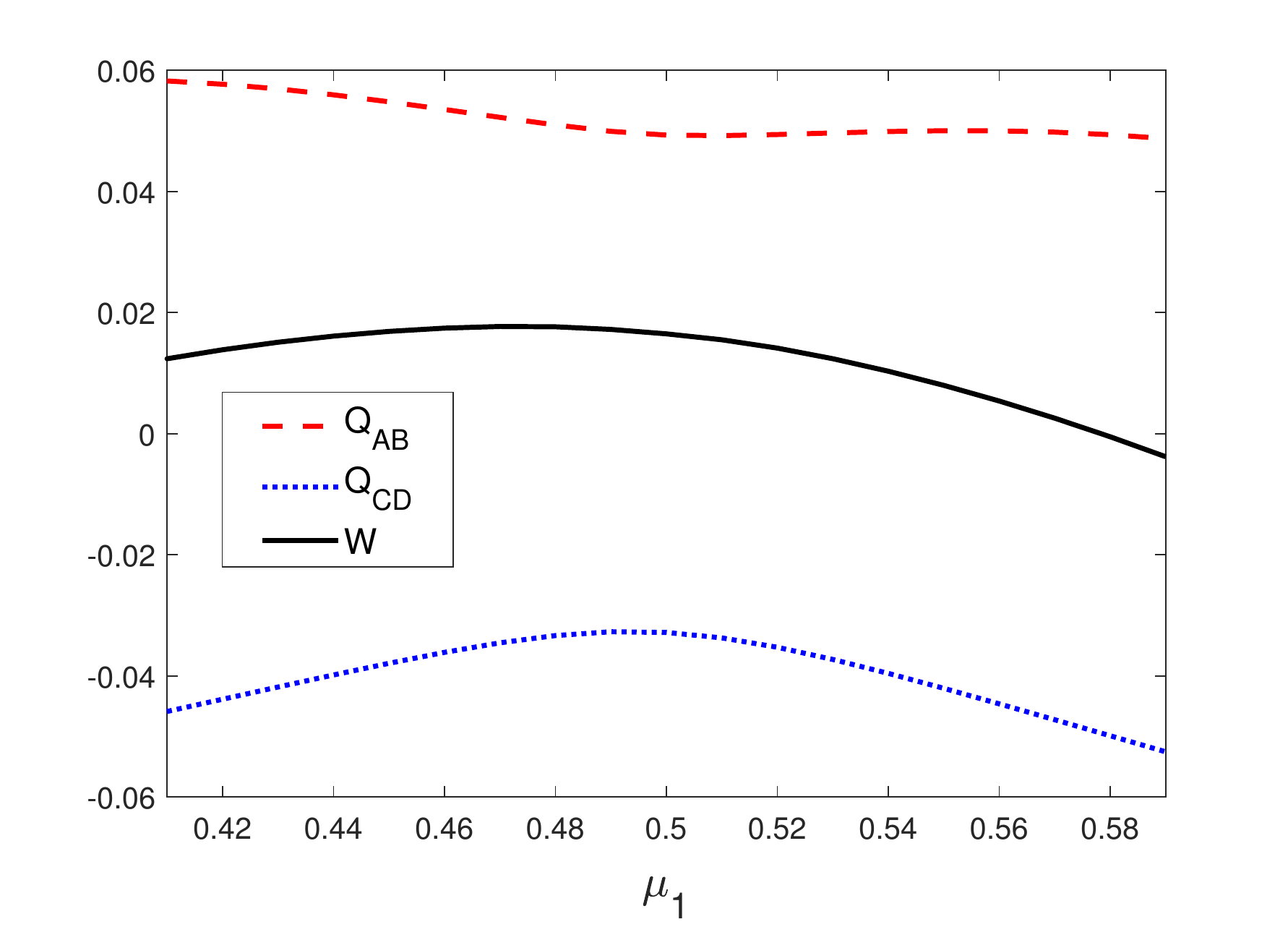}	
	\caption{(Color online)~The absorbed heat $Q_{\text{AB}}$ (red dashed), the ejected heat $Q_{\text{CD}}$ (blue dotted), and the net work output $W$ (black solid) are given as a function of the chemical potential $\mu_1$ of the cold isochore for an ideal Otto cycle working between a hot bath at temperature $T_B= 0.08$ and a cold bath at temperature $T_D = 0.05$.}
	\label{fig:Total1}
\end{figure}

We now fix the parameter of the hot isochore $\mu_2=0.6$ and vary the one for the cold isochore $\mu_1\equiv \mu$ from $0.4$ to $0.6$.
Using the construction of the Otto cycle described in Fig.~\ref{fig:ST1}, we evaluate the work output and the efficiency of the cycle. In Fig.~\ref{fig:Total1}, we plot 
the injected and ejected heat, together with the work output, for a range of $\mu_1$ values. We observe that at the TPT, the  work output of the cycle becomes maximum. The TPT point (or the maximum) is slightly shifted from $\mu_1=0.5$, possibly because the engine operates at finite temperatures, which are known to reduce the critical value of $\mu$ \cite{universality}.  For this range of $\mu_1$, $Q_{\text{AB}}>0$ and $Q_{\text{CD}}<0$, so that the cycle can be properly described as a heat engine operation with $W>0$, except for a small range $\mu_1>0.585$. 

\begin{figure}
	\centering
	\includegraphics[width=8.3 cm,height=5.9cm]{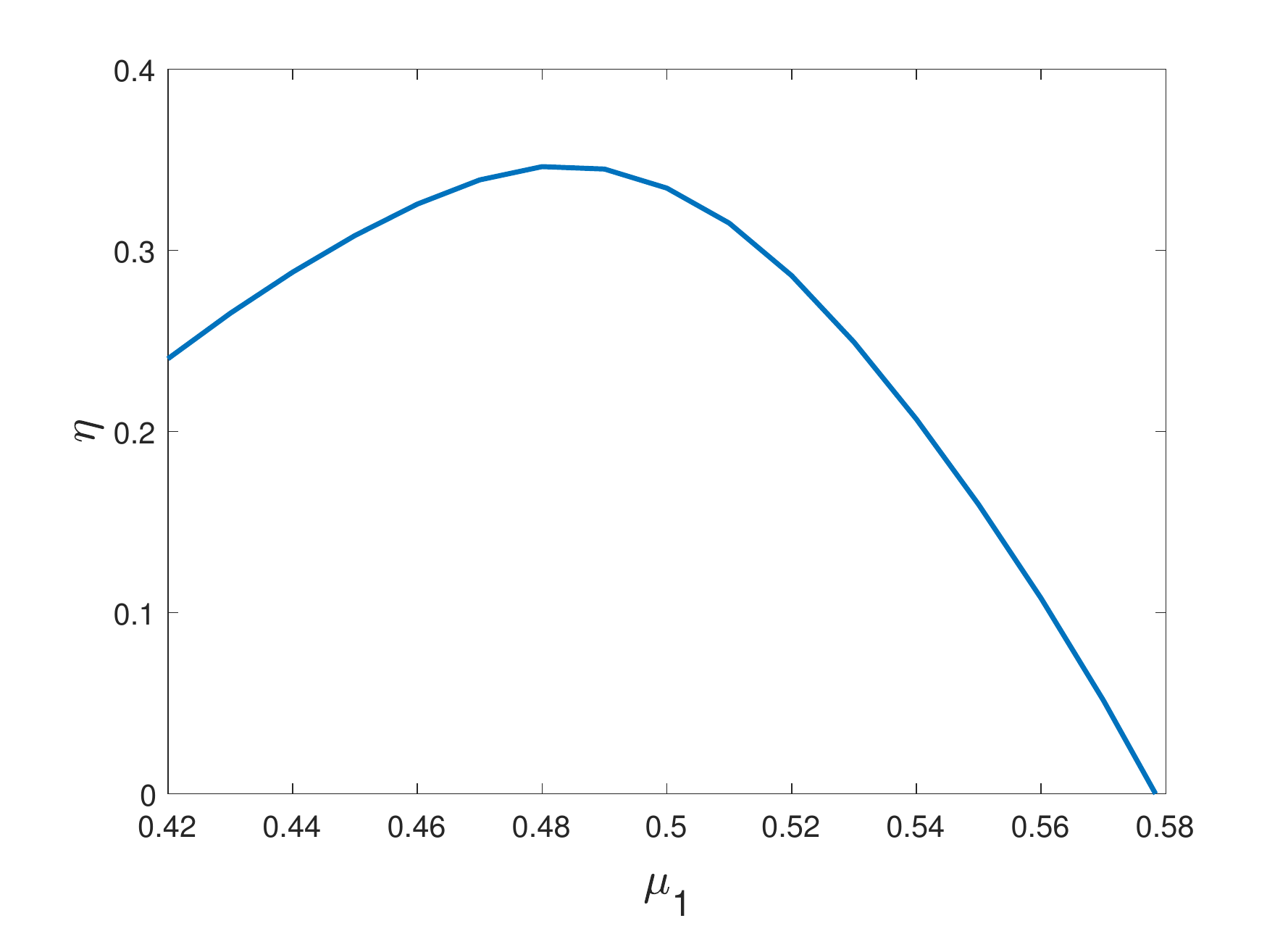}	
	\caption{(Color online)~The efficiency $\eta$ of the finite Kitaev chain heat engine is given as a function of the chemical potential $\mu_1$ of the cold isochore for an Otto cycle working between a hot bath at temperature $T_B= 0.08$ and a cold bath at temperature $T_D = 0.05$.}
	\label{fig:efftotal}
\end{figure}

As shown in Fig.~\ref{fig:efftotal}, the efficiency is maximum at the TPT. Our results show that the TPT enhances  the work output of the Kitaev chain heat engine, as well as its efficiency. Enhancements in the efficiency due to phase transitions have been previously reported in the literature~\cite{Golubeva_PhysRevLett.109.190602}.  The existence of a TPT in a Kitaev chain at finite temperatures is restricted to a certain temperature regime \cite{quelle,delgado1,delgado2}, and the temperatures that we consider here remain within this limit.

\subsection{Bulk and boundary cycles of the Kitaev chain heat engine}\label{sec:bulkboundary}

The Hill's nanothermodynamic framework allows us to examine the bulk and boundary contributions to the net work output of the Kitaev chain heat engine. For that aim, we use Eq.~(\ref{eq:Sbulkedge}) 
to write the total heat exchange
\begin{equation}
Q=\int TdS = \int TdS_c L+ \int TdS_0 = Q_{c}L+Q_{0},
\label{eq:heatbulkboundary} 
\end{equation}
in terms of the bulk and boundary contributions, $Q_c L$ and $Q_0$, respectively.

\begin{figure*}[t!]
	\centering
	\subfloat[]{\includegraphics[width=8.3 cm,height=5.9cm]{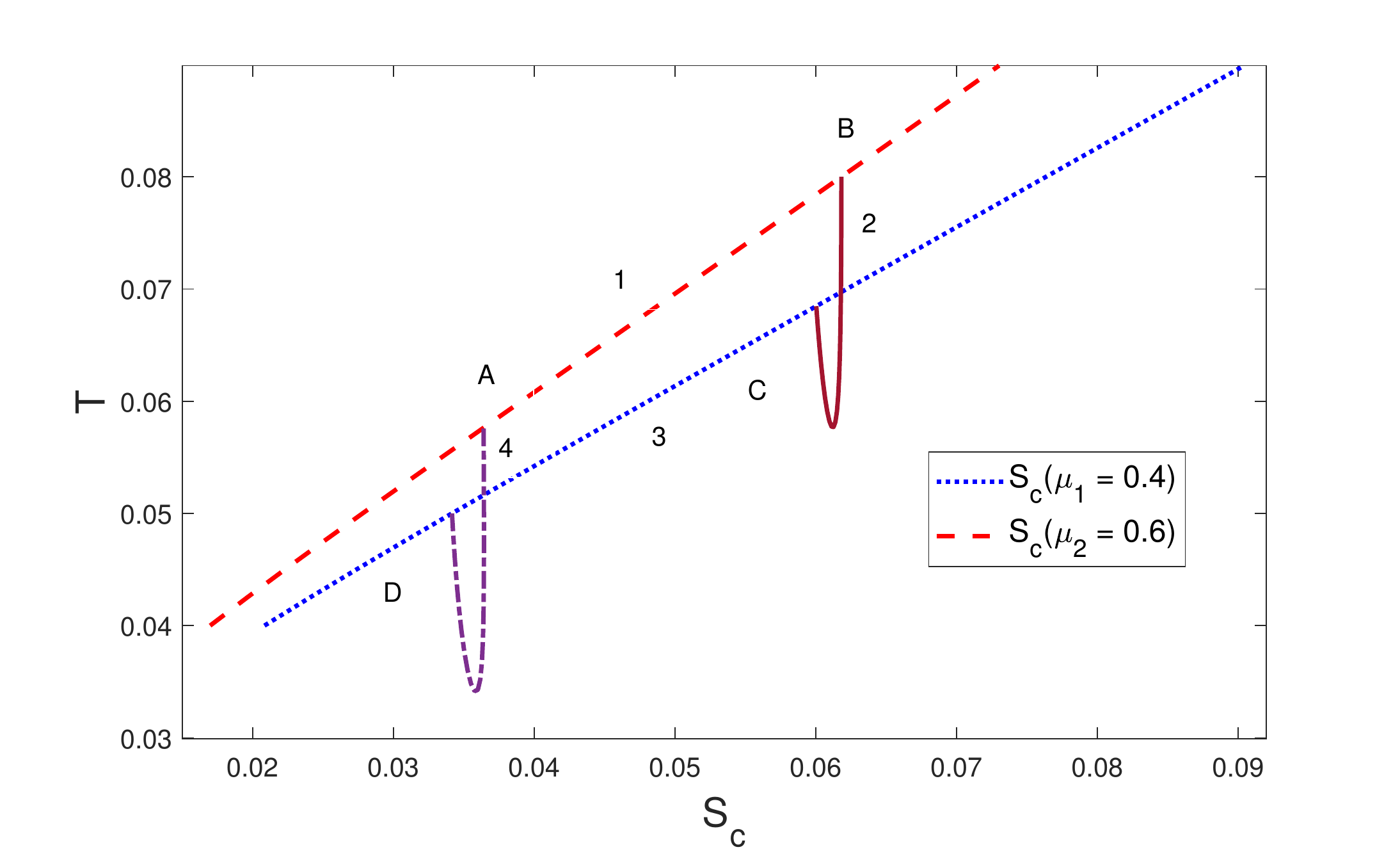}	\label{fig:bulk_TS}}\qquad
	\subfloat[]{\includegraphics[width=8.3 cm,height=5.9cm]{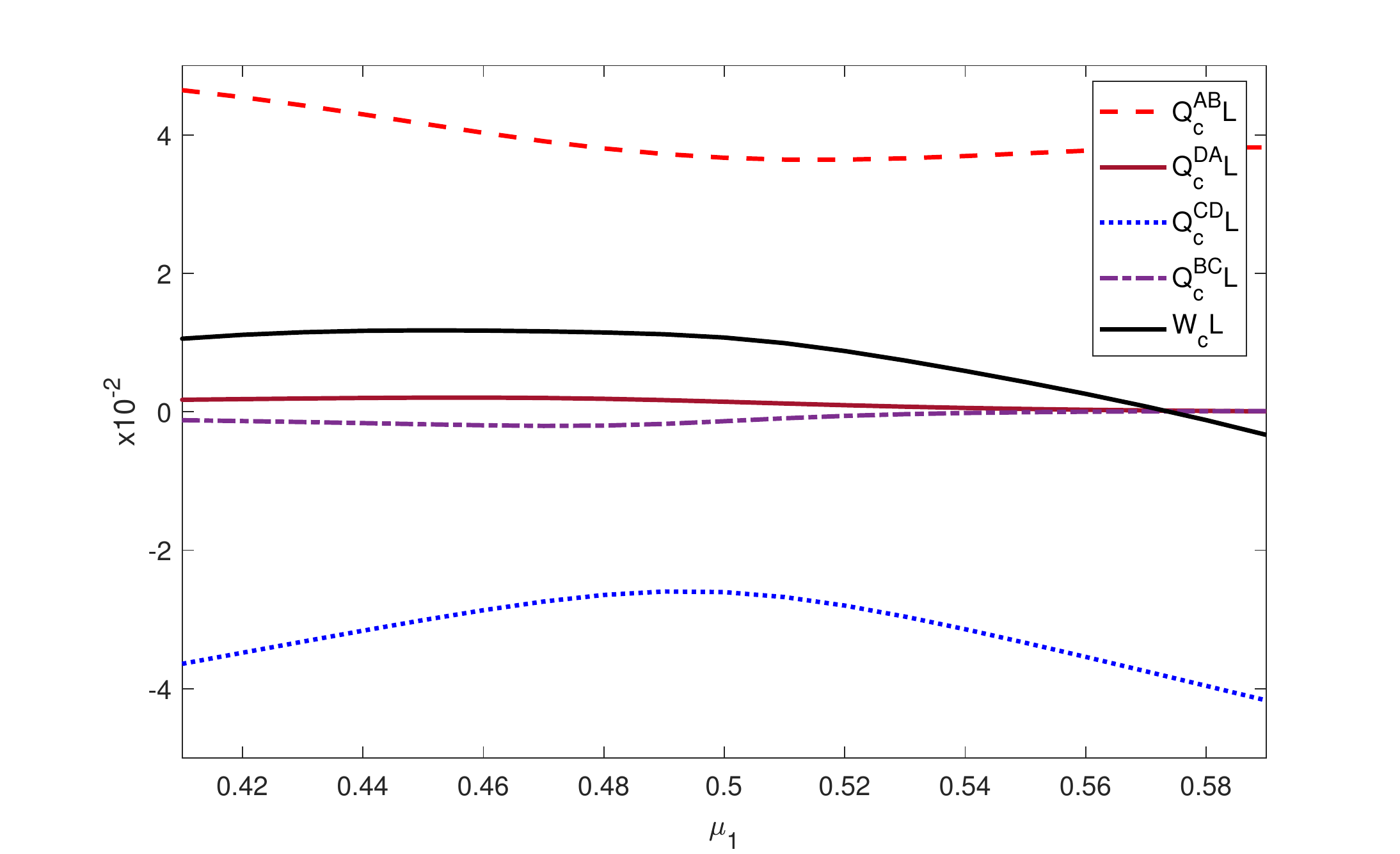}	\label{fig:bulk_heatwork}}
	\caption{(Color Online)~(a) The temperature-entropy density ($T-S_c$) diagramme for the bulk of the finite-length Kitaev chain. The red dashed curve is for the chemical potential $\mu_1=0.4$, while the blue dotted one is for $\mu_2=0.6$. There are heat exchanges on the B-C (brown solid) and D-A (purple dot-dashed) stages.~(b) The heat exchanges of the bulk, $Q_c^{\text{AB}}L$ (red dashed), $Q_c^{\text{DA}}L$ (brown solid), $Q_c^{\text{DC}}L$ (blue dotted), $Q_c^{\text{BC}}L$ (purple dot-dashed), and the bulk work output $W_cL$ (black solid) are given as a function of the chemical potential $\mu_1$ of the cold isochore of a finite-length Kitaev chain in an Otto cycle. The hot isochore chemical potential is $\mu_2=0.6$. All parameters are the same as in Fig.~\ref{fig:ST1}}. 
	\label{fig:bulkcycle}
\end{figure*}

While the total system goes through the Otto cycle described in Fig.~\ref{fig:ST1}, the bulk  goes through the thermal cycle plotted in Fig.~\ref{fig:bulk_TS}. The adiabatic stages are altered due to internal friction resulting from finite-size effects, and the bulk thermal cycle diverges from an ideal Otto cycle. The heat exchanges and the net bulk work output $W_cL$ are given in Fig.~\ref{fig:bulk_heatwork}. We find that the behavior of the heat exchanges by the bulk during the isochoric stages as a function of $\mu_1$ are qualitatively the same as those for the entire chain heat exchanges (cf.~Fig.~\ref{fig:Total1}). 

Although there are heat exchanges during the $B \rightarrow C$ and $D \rightarrow A$ stages in the bulk, given by $Q_c^{\text{BC}}L$ and $Q_c^{\text{DA}}L$ in Fig.~\ref{fig:bulk_heatwork}, their effect on the bulk work output are not strong enough to change the qualitative behaviour of $W_cL.$ We conclude that the bulk thermal cycle is {\it slightly} non-ideal Otto, and the bulk and the total system behave qualitatively similarly in their heat engine operation. The work output of the bulk engine is also maximum at the critical point of the TPT.

 The bulk thermal cycle is only slightly different from the total finite Kitaev chain heat engine cycle and the dependence of the injected and ejected heat and the net work outputs on the chemical potential $\mu_1$ do not exhibit much qualitative difference for the bulk and for the entire chain. However, the boundary behaviour is remarkably different from both.

\begin{figure*}[t!]
	\centering
	\subfloat[]{\includegraphics[width=8.3 cm,height=5.9cm]{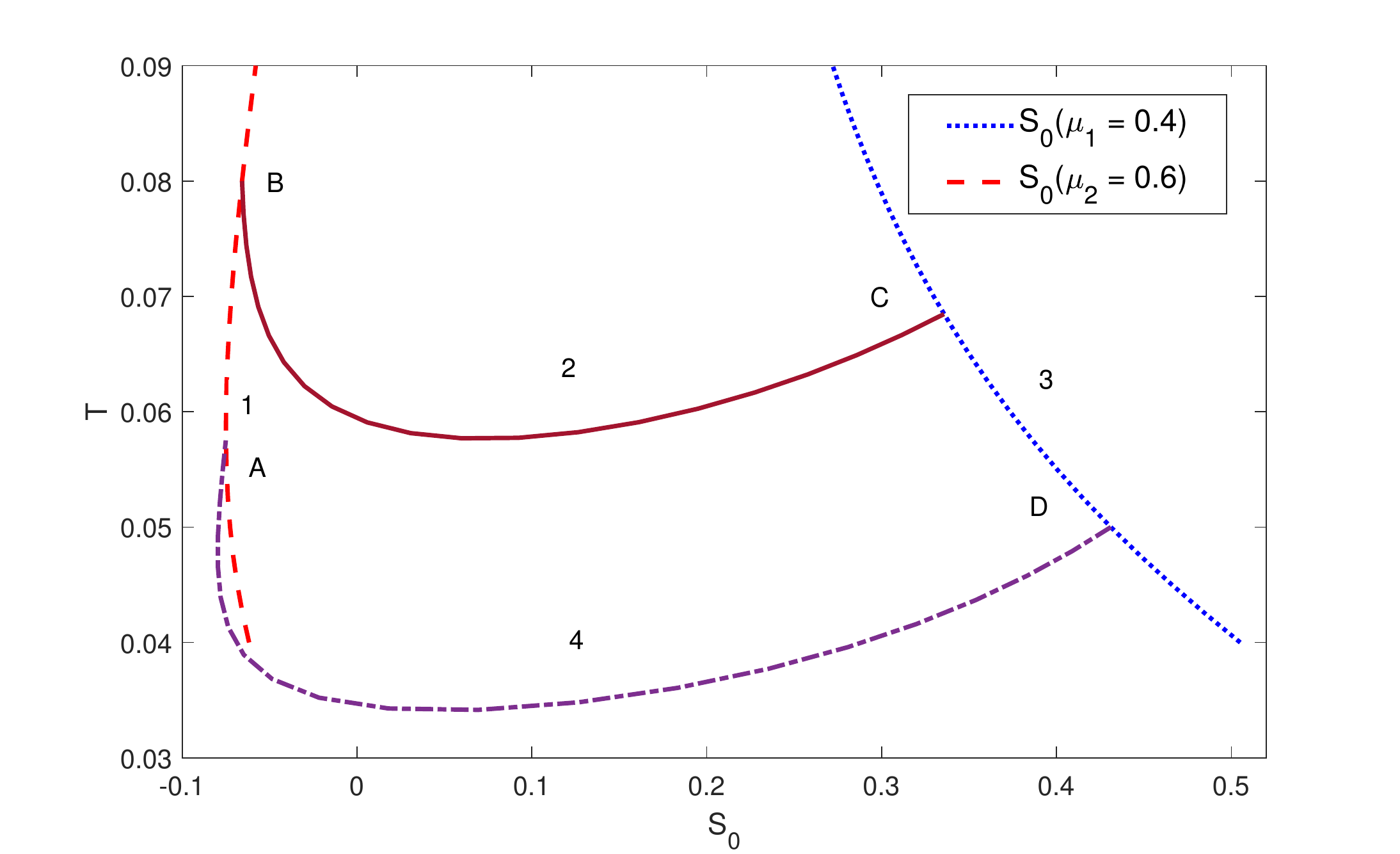}	\label{fig:TSboundary}}\qquad
	\subfloat[]{\includegraphics[width=8.3 cm,height=5.9cm]{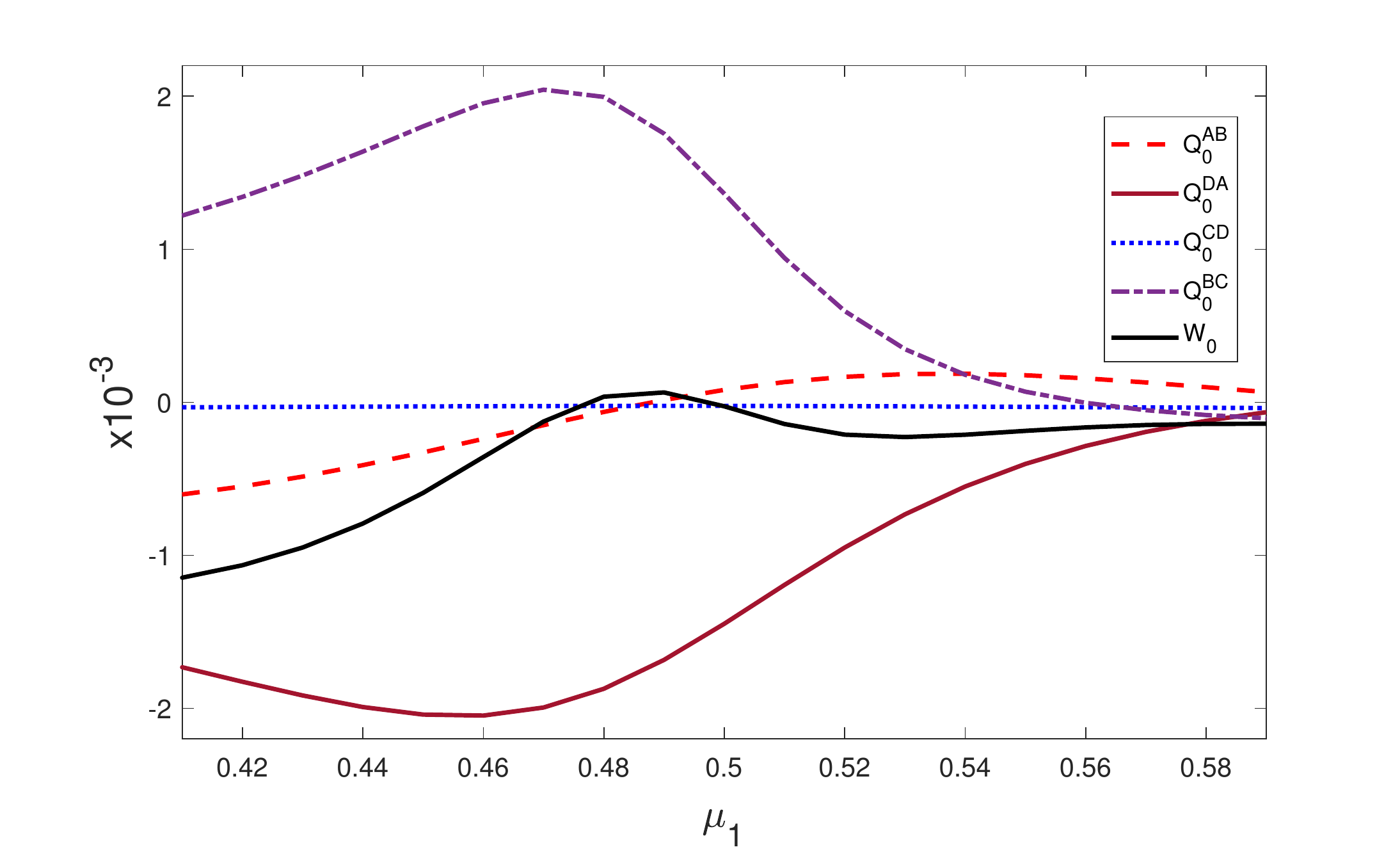}	\label{fig:boundary_heatwork}}
	\caption{(Color Online)~(a)The temperature-entropy ($T-S_0$) diagramme  for the boundary of the finite-length Kitaev chain. The blue dotted curve is for the chemical potential $\mu_1=0.4$, while the red dashed one is for $\mu_2=0.6$. The process from B to C is given by the brown solid line and from D to A by the purple dot-dashed line.  ~(b) The heat exchanges of the boundary, $Q_0^{\text{AB}}$ (red dahed), $Q_0^{\text{DA}}$ (brown solid), $Q_0^{\text{CD}}$ (blue dotted), $Q_0^{\text{BC}}$ (purple dot-dashed), and the boundary work output $W_0$ (black solid) are given as a function of the chemical potential $\mu_1$ of the cold isochore of a finite-length Kitaev chain in an Otto cycle. All parameters are the same as in Fig.~\ref{fig:ST1}.}
	\label{fig:boundary_cycle}
\end{figure*}

 The boundary thermal cycle is plotted in Fig.~\ref{fig:TSboundary}. This cycle produces positive work only for a small parameter range around the TPT point at $\mu_1=0.5,$ where the boundary work, $W_0$ is also maximized. 
 
 Inspecting both Fig.~\ref{fig:bulk_heatwork} and Fig.~\ref{fig:boundary_heatwork}, we observe that $Q_c^{BC}L=-Q_0^{BC}$ (and $Q_c^{DA}L=-Q_0^{DA}$), since there is no heat exchange in the total Kitaev chain heat engine on the adiabatic stage from point B to C (and D to A) and the LHS of Eq.~\ref{eq:heatbulkboundary} would be zero.
 The entropy flows from the bulk to the boundary, consistent with the conservation of total entropy $\Delta S=0$. Local entropy changes in the bulk and boundary are equal and opposite to each other, $\Delta S_cL=-\Delta S_0$. While the adiabatic transformation induced by changing chemical potential is employed on the total system,
 heat transfer from bulk to the boundary causes deviation of ideal adiabatic
 transformations in either subsytem. Accordingly their cycles are non-ideal Otto cycles. This finite-size effect is a spatial analog of the finite-time imperfection of adiabatic transformations~\cite{PhysRevE.65.055102}. Following the term `ìnternal friction'' used for imperfect finite-time adiabatic transformation in finite-time Otto cycles~\cite{PhysRevE.65.055102}, here we dub the finite-size
 effect on adiabatic transformation as ``internal geometric friction''.
 
 The boundary cycle shows different characteristics for the trivial ($\mu_1>0.5$)  and topological ($\mu_1<0.5$)  phases of the finite Kitaev chain. For the topological phase, we can define the incoming heat as $Q_0^{\text{in}}=Q_0^{\text{BC}}+Q_0^{\text{CD}}>0$ and the outgoing heat $Q_0^{\text{out}}=Q_0^{\text{DA}}+Q_0^{\text{AB}}<0.$ Then, the boundary cycle operates as a refrigerator with  $W_0<0.$ In the topological phase, the boundary and bulk thermal cycle operations are reversed. For the trivial phase, the thermal cycle characteristic of the boundary cannot be identified with a heat engine or a refrigerator.
 
Our results presented in this section show that as the total system operates in an ideal Otto cycle, the bulk of the finite Kitaev chain operates {\it approximately close} to an ideal Otto cycle, and the boundary
is a highly peculiar non-ideal Otto cycle. In the appendix~\ref{sec:app}, we also address the question of whether it is possible to identify independently operating bulk, boundary, and total Otto cycles.


\subsection{Landau-Zener and Spin-$1/2$ Model Quantum Heat Engines }\label{sec:LZ}
We show in this section how quantum phase transitions of the Landau-Zener model, both in the case of level crossing and avoided crossing, are probed via a quantum Otto heat engine cycle. We draw an analogy between the bulk/ boundary contributions to the Kitaev heat engine and avoided/level crossing cases of a Landau-Zener model.  
The Landau-Zener model describes a single two-level system, given by the Hamiltonian,
\begin{eqnarray}
H_{\text{LZ}} = -\frac{\omega_0}{2}\sigma_z+g\sigma_x,
\label{HLZ}
\end{eqnarray}
where $\sigma_x$ and $\sigma_z$ are Pauli spin matrices. The gap is determined by  both the parameter $g$ and the transition frequency $\omega_0$.  For the $g = 0$ case, a spin-1/2 model is retrieved. 

\begin{figure}
	\centering
	\includegraphics[width=7.3 cm,height=5.9cm]{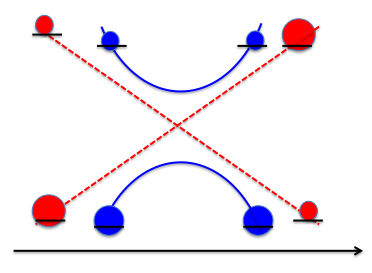}	
	\caption{(Color online)~We display the population inversion in the case of level crossing by red dotted lines. The populations remain in their energy levels in the avoided crossing regime displayed by blue solid lines.}
	\label{fig:LcAcSchematic}
\end{figure}

The ground state energy is $E_0=-\frac{1}{2}\sqrt{4g^2+\omega_0^2}$ and the excited state is given by $E_1=\frac{1}{2}\sqrt{4g^2+\omega_0^2}.$ The energy spectrum  of Eq.~(\ref{HLZ}) displays two different behaviours, depending on the value of the coupling $g$:~(1) For 
$g=0,$ it exhibits  a level crossing of energy levels at the critical value $\omega_0=0,$ and (2) for $g\neq0,$ it shows an avoided crossing at the critical value $\omega_0=0.$  The level crossing gives rise to a first-order quantum phase transition (QPT), while the avoided crossing leads to a second-order QPT~\cite{latency}. A schematic of these two behaviours of the Landau-Zener model at the critical point is given in Fig.~\ref{fig:LcAcSchematic}.

\begin{figure}
	\centering
	\includegraphics[width=7.3 cm,height=5.9cm]{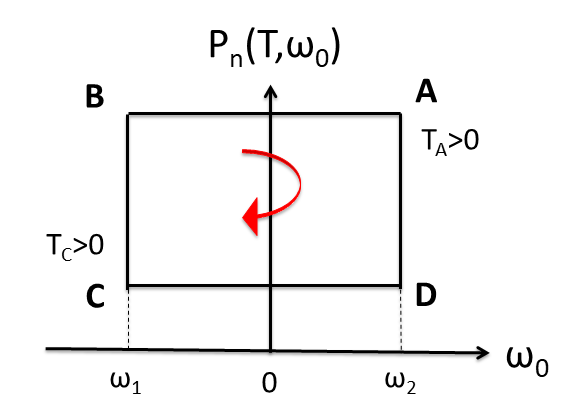}	
	\caption{(Color online)~Quantum Otto engine scheme, i.e. occupation probability-transition frequency $P_n(T,\omega)$ vs $\omega_0$ for both the spin-$1/2$ system and the Landau-Zener model with $g\neq0$ (cycle direction is the same for both cases and is given by the red arrow). The system is in contact with a cold bath at point A and a hot bath at point C.  The cold and hot temperatures are $T_A=0.5$ and $T_C=2,$ respectively. The hot and cold isochore parameter is kept constant at $\omega_0=\omega_1$ and $\omega_0=\omega_2,$ respectively.}
	\label{fig:qcycle}
\end{figure}

Next, we show that it is possible to probe QPTs of the Landau-Zener model for both avoided crossing and level crossing via the work output of quantum heat engine cycles. The heat engine operates using a quantum Otto cycle, consisting of quantum adiabatic and quantum isochoric stages \cite{Quan2007}, as depicted in Fig.~\ref{fig:qcycle}. The control parameter in the cycle is the transition frequency $\omega_0,$ which takes values between the hot isochore frequency $\omega_1$ and the cold isochore frequency $\omega_2.$ 
In the cycle for the spin-1/2 engine, negative temperatures are used to implement population inversion. Consideration of negative temperatures in the context of heat engines and refrigerators to examine the foundations and 
extent of thermodynamical laws is an old problem \cite{Landsberg_1977}, which recently has received much attention due to the recognition of the advantages of negative temperatures in quantum thermodynamics  \cite{Tacchino,PhysRevLett.120.250602,PhysRevLett.122.240602,Xi_2017,Campisi_2016}. The system goes through the following stages for the spin-1/2 engine with $g = 0$ and the Landau-Zener engine with $g = 0.1$ (the cycle direction is the same for the both engines and  is given by the red arrow in Fig.~\ref{fig:qcycle}).
\begin{itemize}
	\item Stage 1 (A to D): The system is attached to a cold bath  with positive temperature at point A, and starts at a thermal state with $\omega_2>0$, $T_A>0.$  The system is detached from the cold bath and isochoric heat transfer takes place at fixed  $\omega_2>0.$ The heat $Q^{\text{in}}_{\alpha}$ is injected to the system:
	\begin{equation}\label{eq:qout12}
	Q^{\text{in}}_{\alpha}=\sum_{n}E_n(\omega_2)[P_n(T_D,\omega_2)-P_n(T_A,\omega_2)].
	\end{equation}	
	Here, $\alpha$ labels either the spin-1/2 system with 1/2 or the Landau-Zener system with LZ. 
	The spin-1/2 system (Landau-Zener system) attains negative (positive) temperature $T_D<0$ ($T_D>0$) at point D.
	\item 	Stage 2 (D to C): The system is quantum adiabatically transformed to a thermal state at point C, where it is attached to a hot bath at temperature $T_C>0.$ The system passes through a level crossing ( an avoided crossing) point at $\omega_0=0$ for the spin-1/2 system (for the Landau-Zener system), as the transition frequency is changed. There is no heat transfer but work is done on the system. A population inversion occurs in the spin-1/2 system as the temperature changes from $T_D<0$ to $T_C>0.$
	\item 	Stage 3 (C to B): The system is separated from the hot bath. An isochoric heat transfer takes place at fixed $\omega_1 < 0.$ The temperature at point B is $T_B<0$ ($T_B>0$) for spin-1/2 system (for the Landau-Zener system). No work is done but heat $Q^{\text{out}}_{\alpha}$ is ejected by the system:
	\begin{equation}\label{eq:qin12}
	Q^{\text{out}}_{\alpha}=\sum_{n}E_n(\omega_1)[P_n(T_B,\omega_1)-P_n(T_C,\omega_1)].
	\end{equation}
	\item 	Stage 4 (B to A): The system is quantum adiabatically transformed to another state at point A passing through the critical point at $\omega_0=0$ as the transition frequency is changed from $\omega_1$ to $\omega_2$. There is no heat transfer but work is done. A population inversion occurs for the spin-1/2 system, as $T_B<0$ at point B and $T_A>0$ at point A. For the Landau-Zener system, $T_B>0,$ and the populations remain in their energy levels.  
\end{itemize}
In  Eq.~(\ref{eq:qin12}) and Eq.~(\ref{eq:qout12}), $$P_n(T_i,\omega_0)=\frac{e^{-\beta_i E_n(\omega_0)}}{Z(T_i,\omega_0)}$$ are the occupation probabilities corresponding to energy $E_n(\omega_0)$ with $n=1,2.$ $i$ labels the cycle points with $i=A, B, C, D.$ $Z(T_i,\omega_0)=\sum_ne^{-\beta_i E_n(\omega_0)}$ is the partition function.  $\beta_i=1/(\kB T_i),$  where $\kB=1$ is the Boltzmann constant.

\begin{figure*}[t!]
	\centering
	\subfloat[]{\includegraphics[width=7 cm,height=5.47cm]{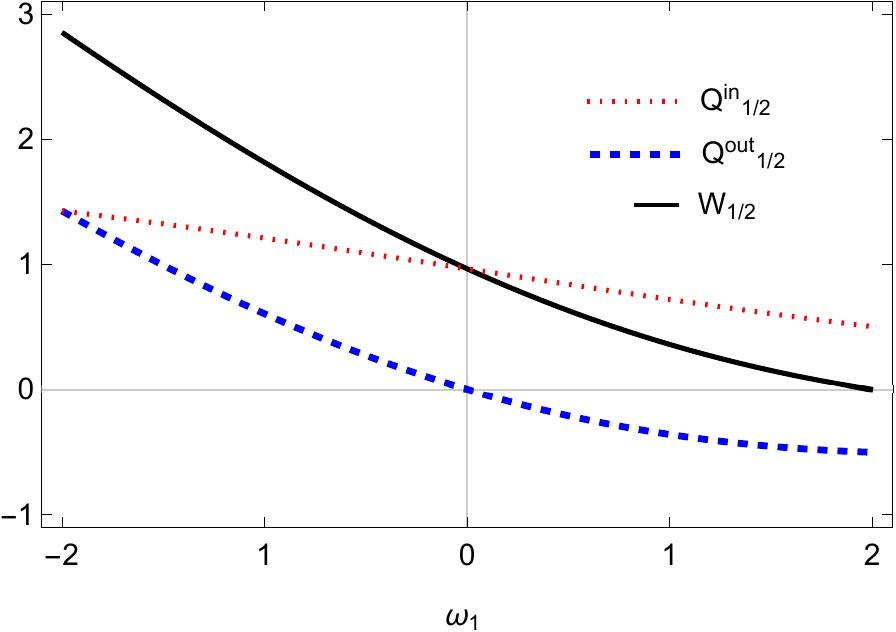}\label{fig:spin1/2}}\qquad
	\subfloat[]{\includegraphics[width=7.5 cm,height=5.47cm]{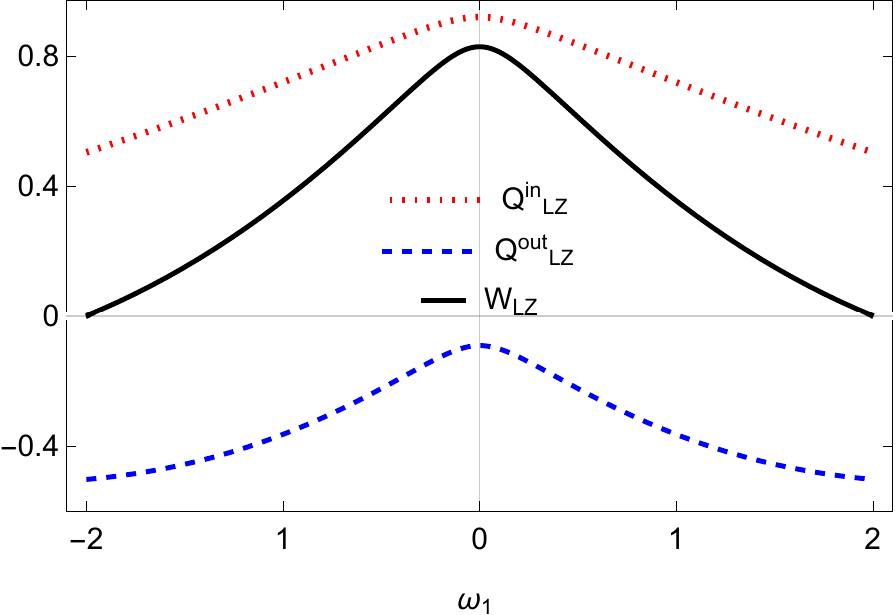}	\label{fig:LZ}}
	\caption{(Color online)~(a) Incoming heat $Q^{\text{in}}_{1/2}$ (red dotted), outgoing heat $Q^{\text{out}}_{1/2}$  (blue dashed) and work output  $W_{1/2}$ (black solid) for a spin-$1/2$ system, (b) incoming heat $Q^{\text{in}}_{\text{LZ}}$ (red dotted), outgoing heat $Q^{\text{out}}_{\text{LZ}}$  (blue dashed) and work output  $W_{\text{LZ}}$ (black solid) for the Landau-Zener model with $g=0.1$, given as a function of  $\omega_1$ of the hot isochore. The initial transition frequency $\omega_0=\omega_2=2$ is held fixed and and $\omega_1$ is changed in the figure.}
	\label{fig:QHE}
\end{figure*}

Fig.~\ref{fig:spin1/2} displays the incoming heat $Q^{\text{in}}_{1/2}$, outgoing heat $Q^{\text{out}}_{1/2}$ and work output  $W_{1/2}$ of the spin-1/2 quantum heat engine working between a cold bath at temperature $T_A=0.5$ and a hot bath at temperature $T_C=2.$ The system is preapared at point A with $\omega_0=\omega_2=2.$
Fig.~\ref{fig:LZ} displays the incoming heat $Q^{\text{in}}_{\text{LZ}}$, outgoing heat $Q^{\text{out}}_{\text{LZ}}$ and work output  $W_{\text{LZ}}$ of the Landau-Zener quantum heat engine in its avoided crossing regime.
\begin{figure*}[t!]
	\centering
	\subfloat[]{\includegraphics[width=8.3 cm,height=5.9cm]{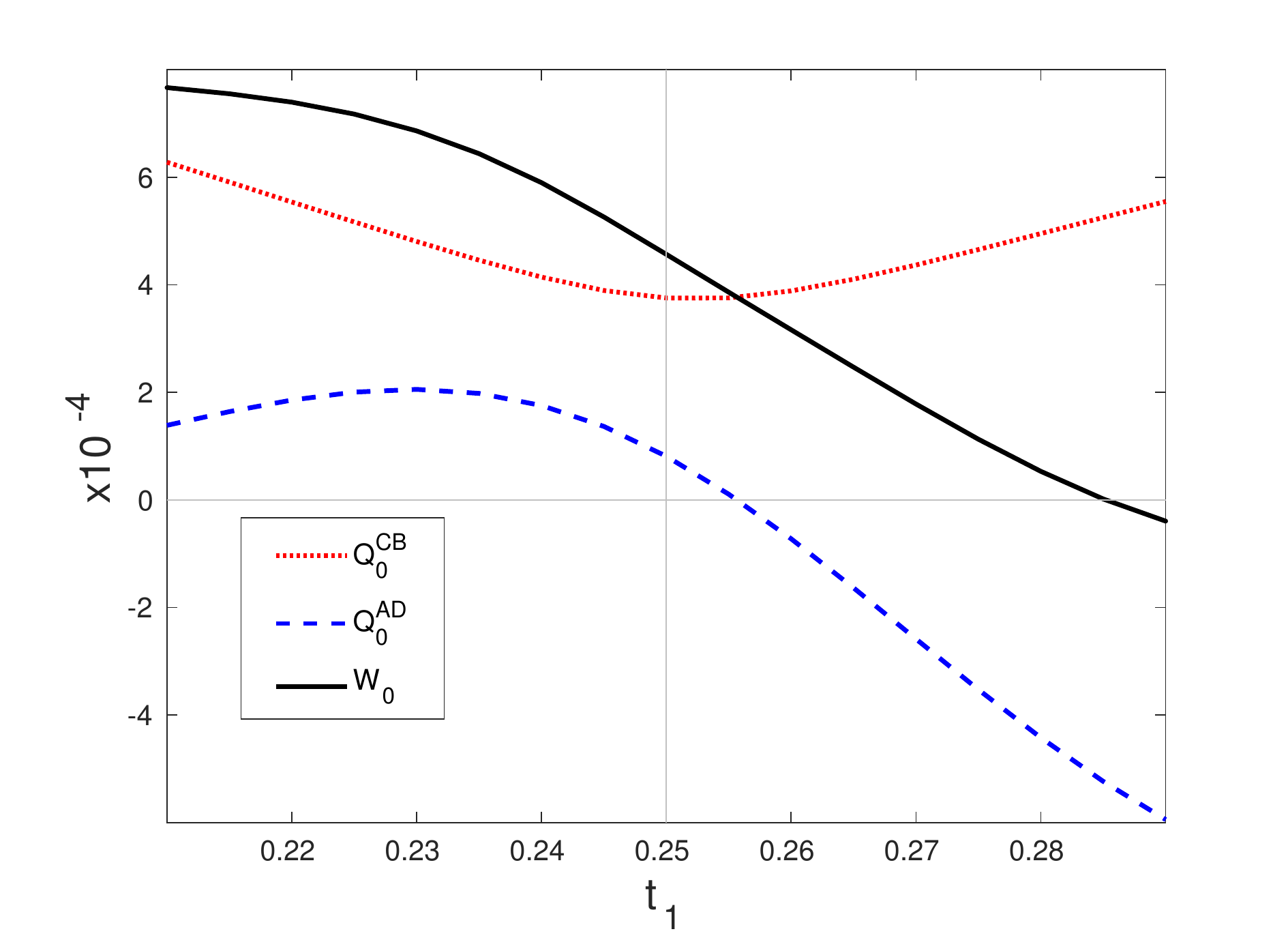}\label{fig:boundarycontr}}\qquad
	\subfloat[]{\includegraphics[width=8.3 cm,height=5.9cm]{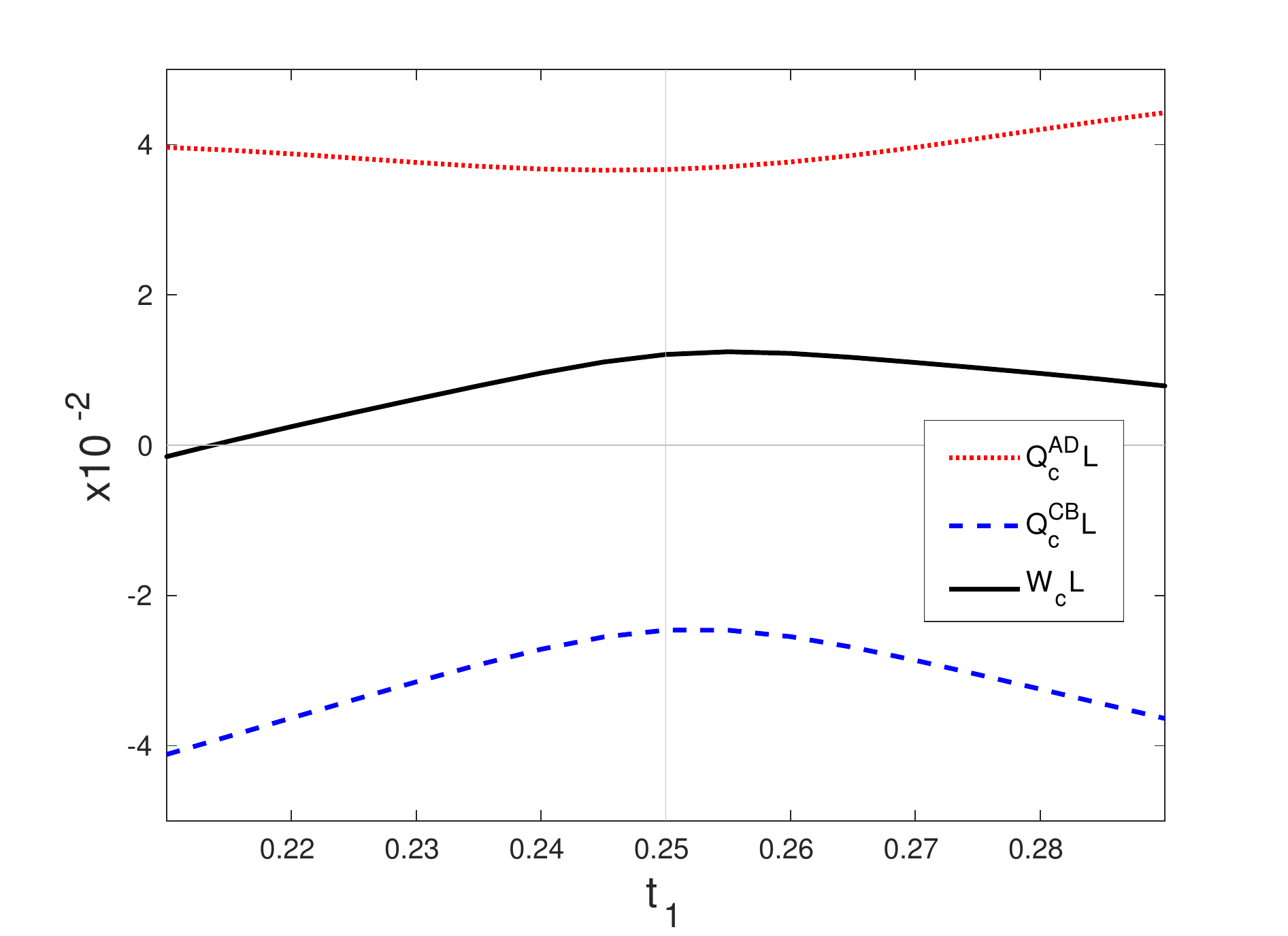}	\label{fig:bulkcontr}}
	\caption{(Color Online)~(a) The boundary and (b) the bulk contributions to the injected $Q^{\text{in}}_0$, $Q^{\text{in}}_c$ (blue dashed) and  ejected heat, $Q^{\text{out}}_0$, $Q^{\text{out}}_c$ (red dotted), and to the work output $W_0$, $W_c$ (black solid), respectively, as a function of the hopping parameter $t_1$ of the first isochore of a finite-length Kitaev chain in an Otto cycle, operating between a cold bath at temperature $T_A= 0.05$ and a hot bath at temperature $T_C = 0.08$. The control parameter of the engine is $t_1$, which changes from 0.2 to 0.3.}
	\label{fig:KHE_t}
\end{figure*}

In Fig~\ref{fig:KHE_t}, we show the boundary and bulk contributions in the Kitaev heat engine, which uses an ideal Otto cycle. We consider a Kitaev chain of length $n=225$ with a superconducting pairing parameter $\Delta=0.25$ as in the rest of the paper with the same cold and hot bath temperatures, $T_A=0.05$ and $T_C=0.08,$ respectively. The chemical potential is fixed at $\mu=0.5$, and the hopping parameter $t$ is used as the control parameter for which a topological phase transition (TPT) takes place at $t=0.25$. For such an ideal Otto cycle, the first isochore parameter $t_1$ is changed, while the parameter of the second isochore $t_2$ is held fixed at 0.3. The cycle direction is the same as in Fig.~\ref{fig:ST1}. We specifically choose the control parameter for the topological engine as the hopping parameter $t$ because its function in the finite-length Kitaev chain Hamiltonian, Eq.~\ref{eq:HKC} is comparable with the function of the transition probability parameter $\omega_0$ in the Landau-Zener Hamiltonian given in Eq.~\ref{HLZ}. 

The behavior for the case of the spin-1/2 model depicted in Fig.\ref{fig:spin1/2} is qualitatively the same as the one found in Fig.~\ref{fig:boundarycontr} for the boundary contribution in the topological Kitaev chain engine. 
The spin-1/2 model captures the first-order quantum phase transition nature of the topological phase transition at the boundary. There, the emergence of unpaired Majorana fermions can be considered as two uncoupled spins for an intuitive description of the heat and work behavior in Fig.~\ref{fig:boundarycontr}.

The intersection of $Q^{\text{in}}_{1/2}$ and $W_{1/2}$ in Fig.\ref{fig:spin1/2} is also a common feature in the heat and the work behaviour of spin-1/2  and the boundary of Kitaev chain systems , which occurs due to the fact that $Q^{\text{out}}_{1/2}=0$. No heat transfer is required to equilibrate the system with the bath at $T_C$, as both levels are degenerate at the transition point ($\omega_0=0$ in Fig. \ref{fig:spin1/2} and $t_1=0.25$ in Fig.~\ref{fig:boundarycontr}). Another common feature is that the thermal cycle characteristics before and after the phase transition are different both in Fig.~\ref{fig:spin1/2} and in Fig.~\ref{fig:boundarycontr}: Before their respective transition points, their thermal cycles cannot be identified with a heat engine or a heat pump, but after the phase transition point, they both operate as heat engines producing positive work.

The Landau-Zener engine also operates according to the scheme given in Fig.~\ref{fig:qcycle}. We note that the temperatures at $T_B$ and $T_D$ are positive for such an engine, since  the system is taken through an avoided crossing point at $\omega_0=0$. The assignment of negative temperatures is used only when there is a population inversion in the system.

The behavior of Fig.~\ref{fig:bulkcontr}, which is the bulk contribution to the topological Kitaev heat engine, is similar to the Landau-Zener quantum engine results depicted in Fig.\ref{fig:LZ}. Both results are consistent with the second-order  nature of the  phase transitions. The work output is maximized at the critical point, $\omega_1=0$ for the Landau Zener quantum heat engine and $t_1=0.25$ for the bulk contribution of the topological finite Kitaev heat engine. Before and after their respective critical points, the thermal cycles can be properly identified as heat engines. 

Thus, we report that thermal cycles display universal features in their heat exchange and work output behaviours according to the order of phase transition taking place.

We remark that enhancement of the work output at the critical point of the phase transition cannot be used per se to justify the topological character of the phase transition. It is a general phenomenon that can emerge in the general class of gap closing quantum phase transitions. It is necessary to examine the simultaneous contributions from the bulk and boundary to the total work. We find that the contribution of the boundary to the work output shows no enhancement, due to a level crossing at the critical point; on the other hand, the avoided crossing leads to an enhancement in the contribution from the bulk to the work output.  

The observations are in parallel with the conclusion that the subdivision potential describes the edge behavior of the system and captures its unique thermodynamic signatures~\cite{quelle}. 
This model-independent conclusion applies both to the Kitaev chain and to the toy (Landau-Zener) Otto engine models.  The work output of the Landau-Zener engine with (without) level crossing exhibits a similar behavior with the edge (bulk) contributions of the Kitaev chain engine.  This provides physical intuition to recognize the topological phase transition in the Kitaev chain at finite temperatures. Physically, the differences in the work output from the bulk and edge of a Kitaev chain engine can be related to differences in the corresponding specific heat.  We determine the order of the phase transitions in the Kitaev chain and in the Landau-Zener model following the Ehrenfest classification. The thermodynamic phase transition can be identified with the topological phase transition at sufficiently low temperatures, determined by the Uhlmann phase~\cite{delgado1,delgado2}. Hence, we conclude that the bulk and edge work output of the Kitaev chain allows one to identify the topological phase transition.


\subsection{Effective Work}\label{sec:effectivework}
Hill’s thermodynamics is traditionally developed for an idealized description of a nanothermodynamics system. A gas of multiple copies of a small system is no longer finite size, but a macroscopic object on which laws of thermodynamics can be applied. With the help of the subdivision potential, the thermodynamic behavior of a single copy can be obtained. For our example of Kitaev chain, this would correspond to a chain immersed in a heat bath, with every site accessing the bath as a whole. On the other hand, one could imagine an infinitely long chain, on which laws of thermodynamics could be applied. The chain is envisioned as a sum of the actual finite-size system plus a semi-infinite chain, so that the extended total chain is macroscopic. In a typical thermal experiment,  the heat bath would be able to access only the sites of the semi-infinite (lead) chain, while the finite-size system component remains inaccessible. Thermalization of the lead chain brings its sites to canonical Gibbs thermal states. Tracing them out makes the Hamiltonian, and hence the energy levels of the finite-size actual system to be temperature dependent. Then, the picture becomes similar to the one of Hill in which a finite-size system is coupled to a heat bath, and the energy levels are temperature dependent. In a nutshell, the difference between thermal properties, such as heat or work transfer, between the two pictures (Hill’s and temperature dependent energy levels, TDELs) arise due to the introduction of an interface between the heat bath and the finite-size system in the case of TDEL. Accordingly, TDEL would be a more faithful representation of a thermal measurement or process in typical experiments using interfaces to access finite-size systems.

For the TDEL approach, we consider a macroscopic Kitaev chain decomposed into two parts: One part is taken as the finite-length system, and the other is an interface lead, which is long enough to be assumed in thermal equilibrium with a heat bath~\cite{Elcock_1957}. TDELs arise as a result of averaging over the interface microstates, and describe the heat exchange during the thermalization of the finite-length chain with the heat bath in a thermodynamically consistent way. The TDELs method, different from Hill's approach, describes the boundary as an energy channel interfacing the bath and the system.

The connection between Hill's nanothermodynamics and TDELs is sustained by recognizing that the subdivision potential acts as a thermal perturbation by the environment~\cite{miguel1,miguel2,miguel_temperature-dependent_2015}. As explained in Sec.~\ref{sec:tdel}, the subdivision potential of the Hill's nanothermodynamics can be used to calculate the heat dissipated through this boundary energy channel, upon associating $X=\Phi_0$ in Eq.~(\ref{eq:Qeff})~\cite{miguel1}. Thus, we use these two nanothermodynamic approaches in conjunction with each other.

\begin{figure*}[t!]
	\centering
	\subfloat[]{\includegraphics[width=8.3cm,height=5.9cm]{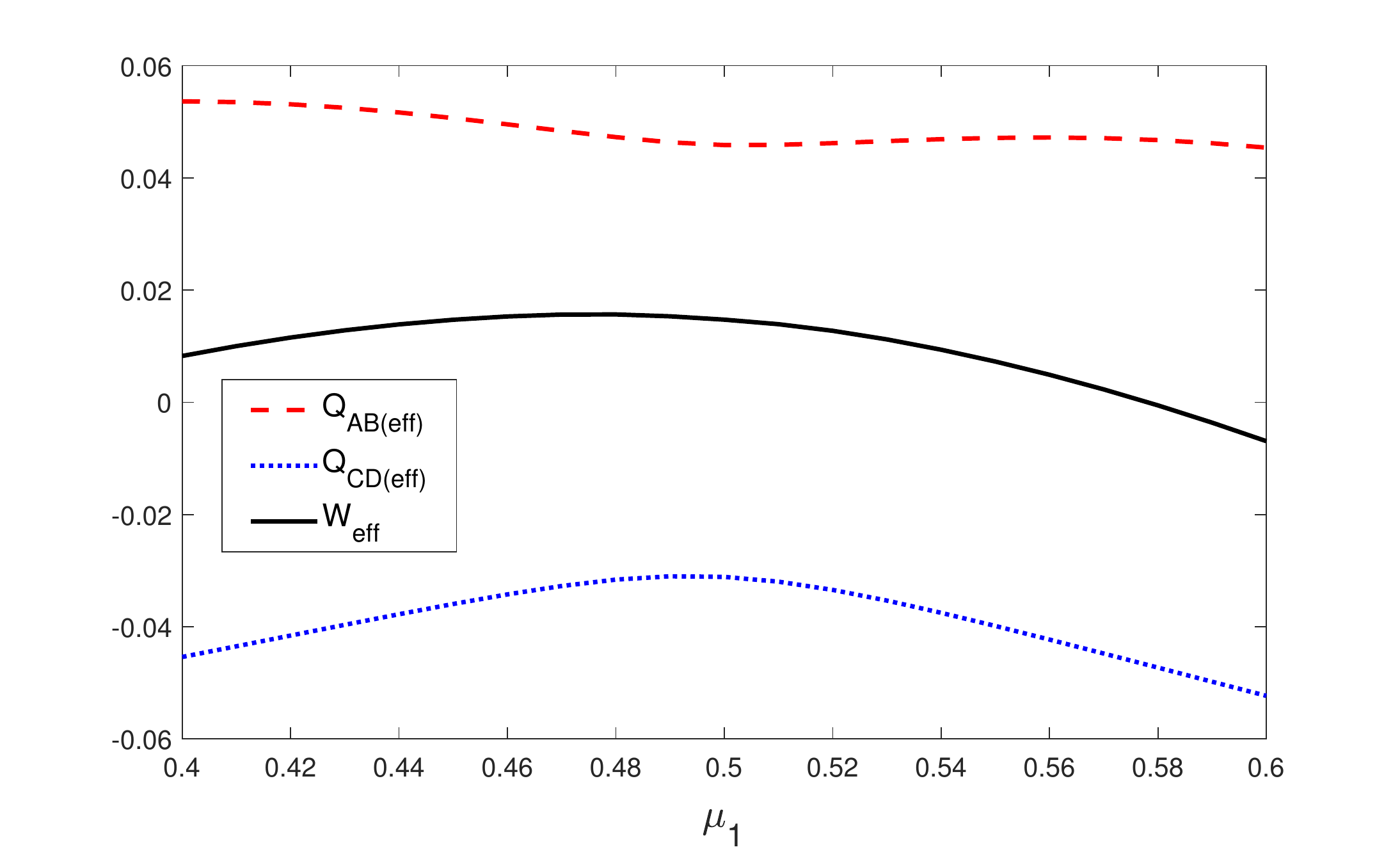}	    \label{fig:figure8a}}\qquad
	\subfloat[]{\includegraphics[width=8.3cm,height=5.9cm]{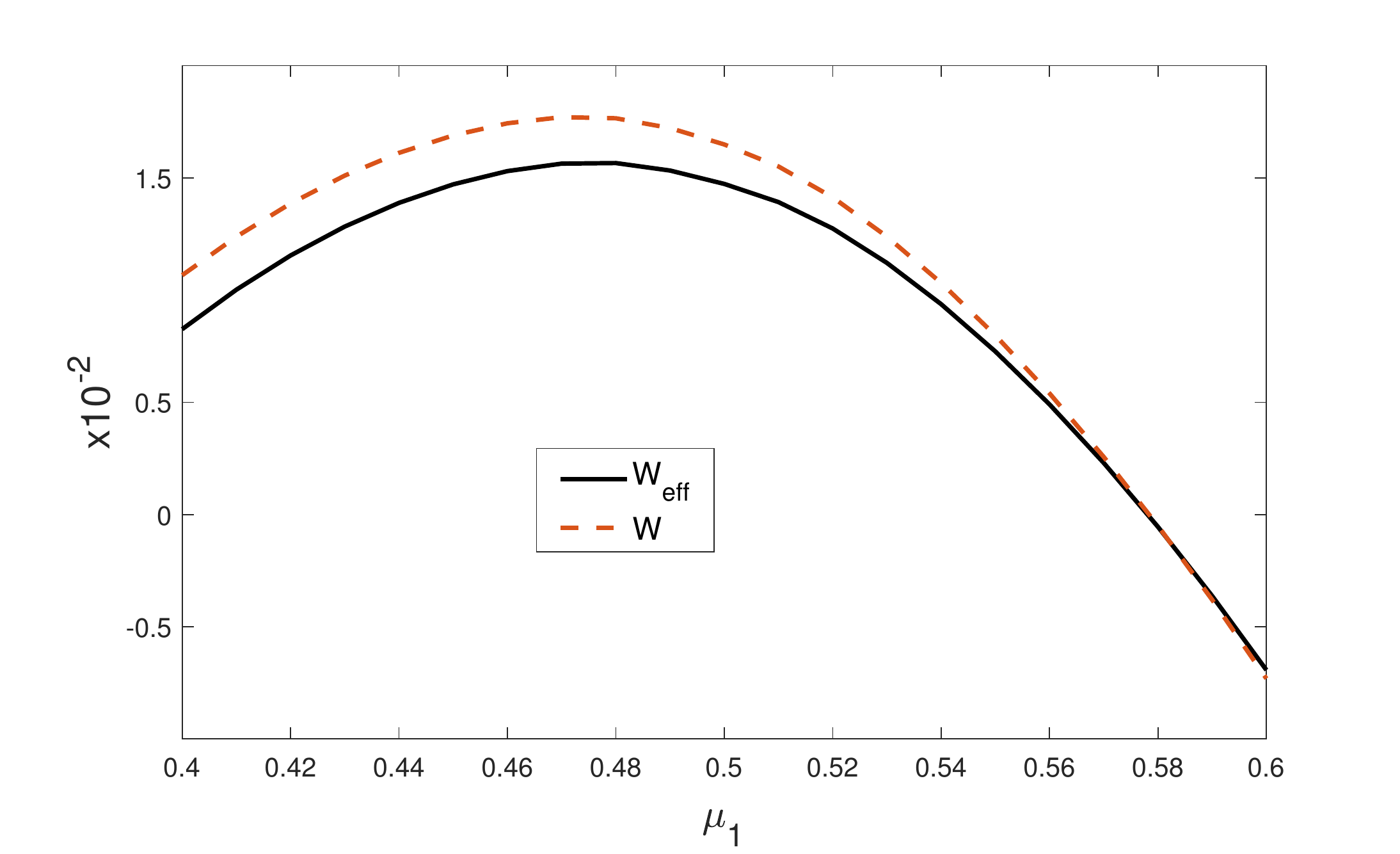}	    \label{fig:figure8b}}
	\caption{(Color online)~ (a)~Effective incoming heat $Q_{\text{AB(eff)}}$ (red dashed), effective outgoing heat $Q_{\text{CD(eff)}}$ (blue dotted)  and effective work $W_{\text{eff}}$ (black solid). (b) Total net work  $W$ (red dashed) and effective work $W_{\text{eff}}$ (black solid) are given as a function of $\mu_1$ for an Otto cycle working between a hot bath at temperature $T_B= 0.08$ and a cold bath at temperature $T_D= 0.05$.}
	\label{fig:figure8}
\end{figure*}

The effective incoming and outgoing  heats are computed, respectively, according to the following formulae:
\begin{eqnarray}
Q_{\text{AB(eff)}}= \int_{S_A}^{S_B}TdS-\int_{T_A}^{T_B}\frac{\Phi_0}{T}dT,
\label{eq:qineff}
\end{eqnarray} 

\begin{eqnarray}
Q_{\text{CD(eff)}}= \int_{S_C}^{S_D}TdS-\int_{T_C}^{T_D}\frac{\Phi_0}{T}dT.
\label{eq:qouteff}
\end{eqnarray} 
The second terms in Eqs.(\ref{eq:qineff}-\ref{eq:qouteff}) are the correction terms, which are computed using $\Phi_0$ of Hill's nanothermodynamics.
The total effective work is
\begin{eqnarray}
W_{\text{eff}}=Q_{\text{AB(eff)}}+Q_{\text{CD(eff)}}.
\label{eq:workeff}
\end{eqnarray} 
We calculate the injected and ejected effective heat and also the effective work for the Otto cycle depicted in Fig.~\ref{fig:ST1} using Eqs.~(\ref{eq:qineff})-(\ref{eq:workeff}). The results are plotted in Fig.~\ref{fig:figure8a}.

\begin{figure*}[t!]
	\centering
	\subfloat[]{\includegraphics[width=8.3cm,height=5.9cm]{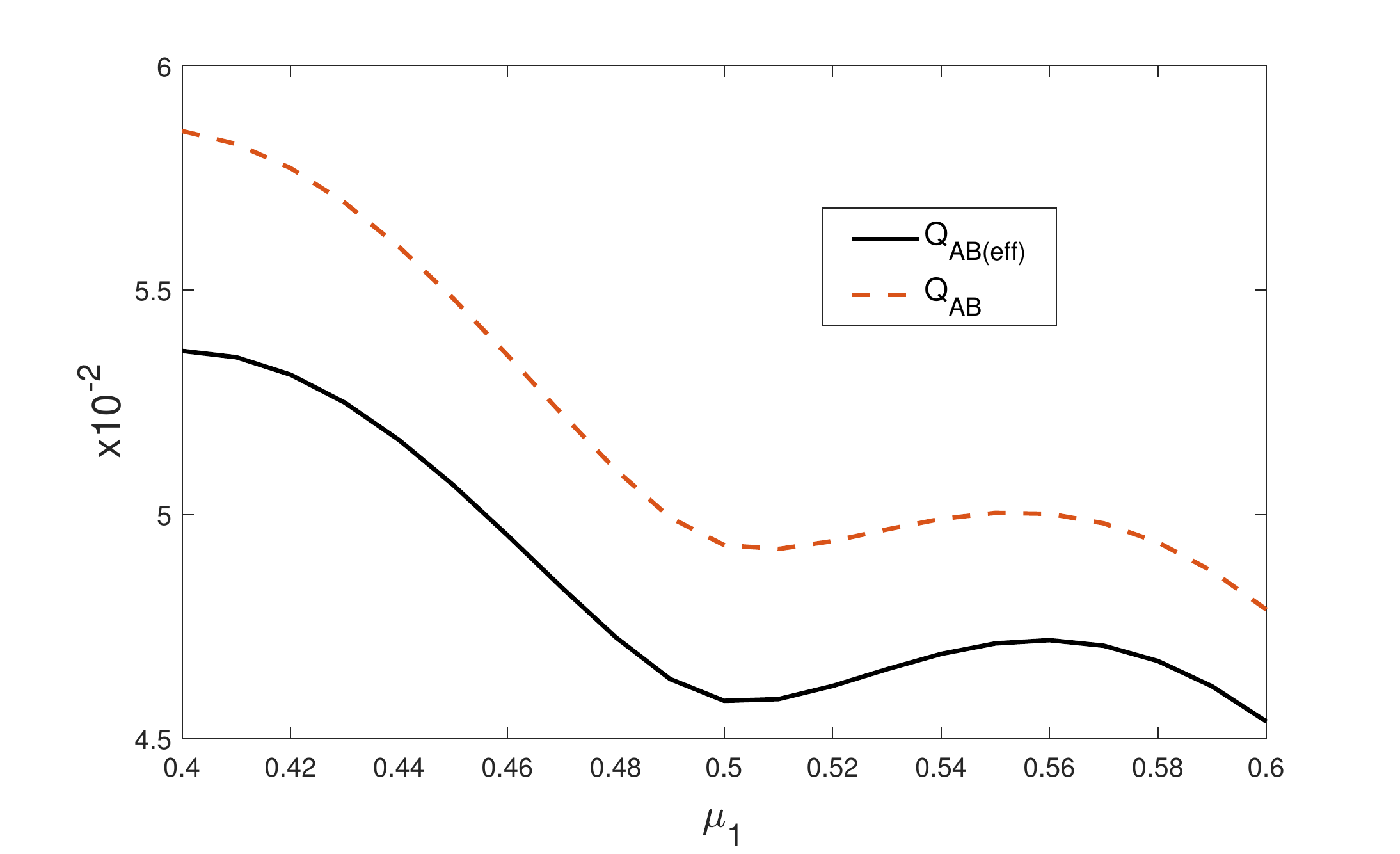}	    \label{fig:figure9a}}\qquad
	\subfloat[]{\includegraphics[width=8.3cm,height=5.9cm]{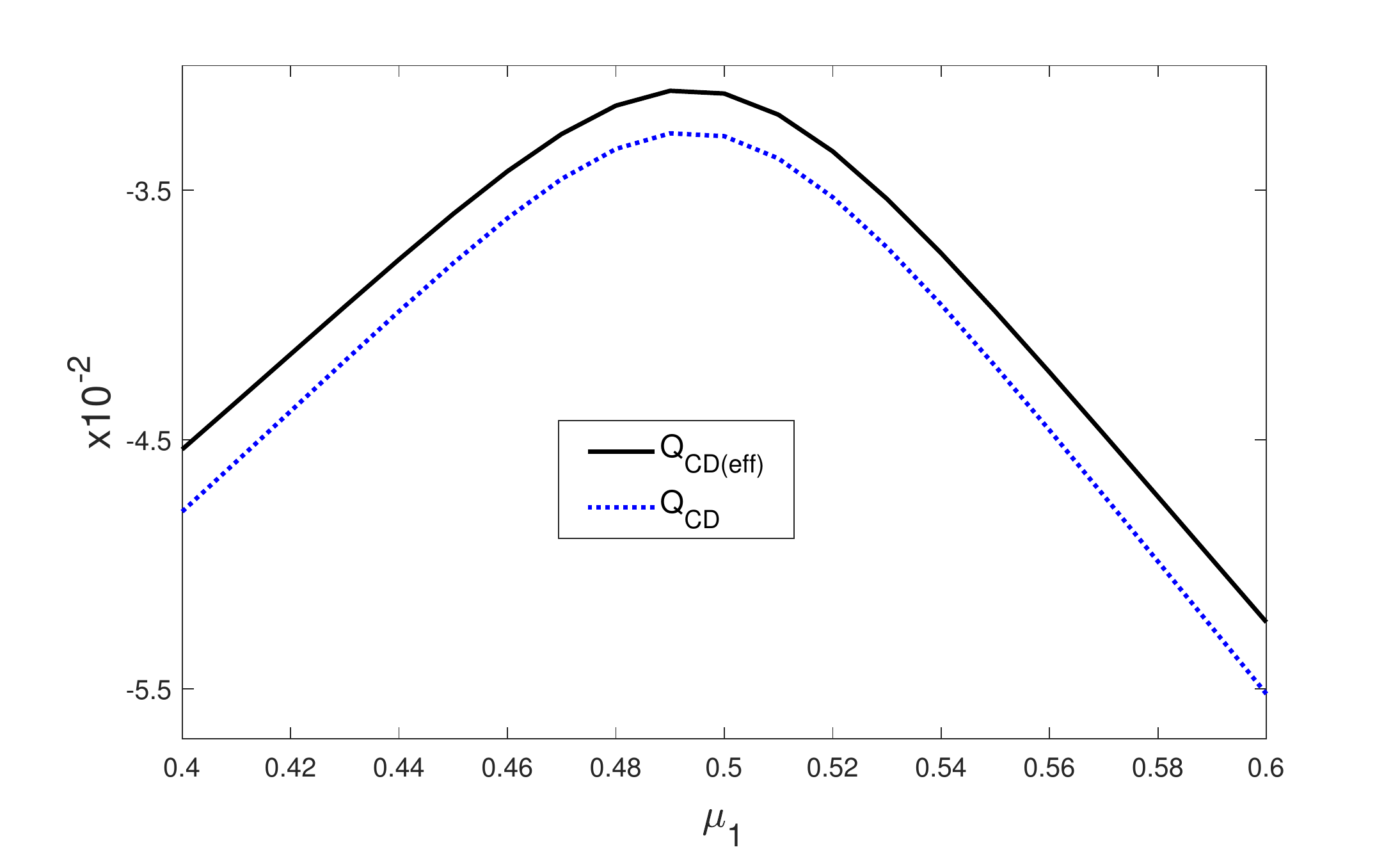}	    \label{fig:figure9b}}
	\caption{(Color online)~ (a) Absorbed heat $Q_{\text{AB}}$ (red dashed) and effective absorbed heat $Q_{\text{AB(eff)}}$ (black solid). (b)~ Ejected heat $Q_{\text{CD}}$ (blue dotted) and effective ejected heat $Q_{\text{CD(eff)}}$ (black solid)
		are given as a function of $\mu_1$ for an Otto cycle working between a hot bath at temperature $T_B= 0.08$ and a cold bath at temperature $T_D = 0.05$.}
	\label{fig:figure9}
\end{figure*}

There is a signature of a TPT in both $Q_{\text{AB(eff)}}$, $Q_{\text{CD(eff)}}$ and in the effective work $W_{\text{eff}}$ at $\mu=2t$. We also present a comparison between the work calculated using Hill's nanothermodynamics and the effective work, which accounts for the TDELs in Fig.~\ref{fig:figure8b}. Both the work and the effective work have qualitatively the same behavior, but the effective work  $W_{\text{eff}}$ acquires lower values. 
 The behavior of the exchanged heat values yielded by Hill's nanothermodynamics are compared with the effective heat values in Fig.~\ref{fig:figure9}.
We observe that $Q_{\text{AB(eff)}}$ is less than $Q_{\text{AB}}$  and $Q_{\text{CD(eff)}}$ is higher than $Q_{\text{CD}}.$ This shows that the heat exchanges are modified by the interface effects taken into account through the TDELs. We note that the qualitative behaviour of heat exchanges and work outputs are the same in both approaches and that there is only a small quantitive difference.
\section{conclusions}\label{sec:conc}
We propose a heat engine, which uses the finite Kitaev chain as its working substance. The heat engine is based on an Otto cycle, where there are two isentropic (adiabatic) and two isochoric (isoparametric) stages. The control parameter in the engine is the chemical potential of the Kitaev chain. We work in the low temperature regime, such that the temperatures are smaller than the energy gap of the system.

We report that the TPT of the finite Kitaev chain enhances both the total work output and the efficiency. We further investigate the thermodynamic properties of the finite-length Kitaev chain within two thermodynamic frameworks for finite-size systems: Hill's nanothermodynamics and TDELs scheme. 

Based on Hill's nanothermodynamics, we identify the qualitative and quantitative features arising from the bulk and boundary in the work output of the Otto cycle. We find that bulk and boundary undergo their own thermal cycles different from the ideal Otto cycle. The bulk operates in a non-ideal Otto heat engine in both phases of the Kitaev chain with a behaviour that is qualitatively similar to the one of the total system.  The boundary cycle yields a positive work output only in a small parameter region around the TPT. The boundary shows two different thermal cycle characters before and after the transition point, associated with its 
first-order TPT behavior. The thermal cycle of the boundary contribution can be identified with neither a heat engine nor a refrigerator in the trivial phase  but acts as a refrigerator in the topological phase of the finite Kitaev chain.

We conclude that the non-ideal Otto cycle behavior, which arises at the bulk and boundary is a finite-size effect, a spatial analog of the finite time internal friction (or quantum friction)~\cite{PhysRevE.65.055102}. Both finite-time and finite-size effects cause imperfect adiabatic transformation, leading to non-ideal Otto cycles.

We also show the existence of three Otto cycles working independently between two baths and making use of the total, bulk and boundary contribution separately. This scheme only works in the {\it topological} phase of the Kitaev chain. While the total system and the bulk operate as heat engines, the boundary operates as a {\it heat pump}. We remark that these cycles attain different intermediate temperatures and cannot be designed to operate simultaneously. Nevertheless, the Kitaev chain can be used as a multi-functional thermal device in which engine and refrigerator operations are tunable and spatially separated to the bulk and to the boundary.  

Using the TDELs scheme in conjunction with the Hill nanothermodynamics approach, we have, in addition, calculated the effective heat exchange and the effective work of the finite Kitaev chain engine. We find that the effective work is qualitatively lower than what is calculated by Hill's nanothermodynamics. The incoming heat is reduced and the outgoing heat is increased  due to the energy dissipation in the system-bath interface. While Hill's nanothermodynamics allows for a clear identification of the bulk and boundary contributions in the heat engine operation, TDELs provides a proper assignment of the heat exchange between the heat baths and the finite system. We show that these two approaches complement each other and can be used in conjunction for detailed modeling of thermal machines and experiments with topological finite systems.
 
We show that, by separating the bulk and boundary contributions to the work output of the Kitaev heat engine, we can identify the order of the phase transition taking place by inspecting these contributions. The boundary exhibits a first-order phase transition, analogous to the level crossing regime of a two-level Landau-Zener model, and the bulk exhibits a second-order phase transition like in the avoided crossing regime of the Landau-Zener model. We state a reservation in this analogy which we leave for further study, which is that the bulk and boundary of the Kitaev heat engine go through their own thermodynamic cycles, which are non-ideal Otto cycles, but still exhibit phase transition features. We note that this may be due to the universal feature of phase transitions and conclude that those seem to be independent of the cycle being considered. 

\section*{Acknowledgements}
We acknowledge support by the Scientific and Technological Research Council of Turkey (T\"{U}B{\.I}TAK), 
Grant No.~ (117F097) and by the EU-COST Action (CA16221). The authors would like to thank S.N. Kempkes and R. Arouca for a careful reading of the manuscript.

\begin{figure*}[htb!]
	\centering
	\subfloat[]{\includegraphics[width=8.3cm,height=5.9cm]{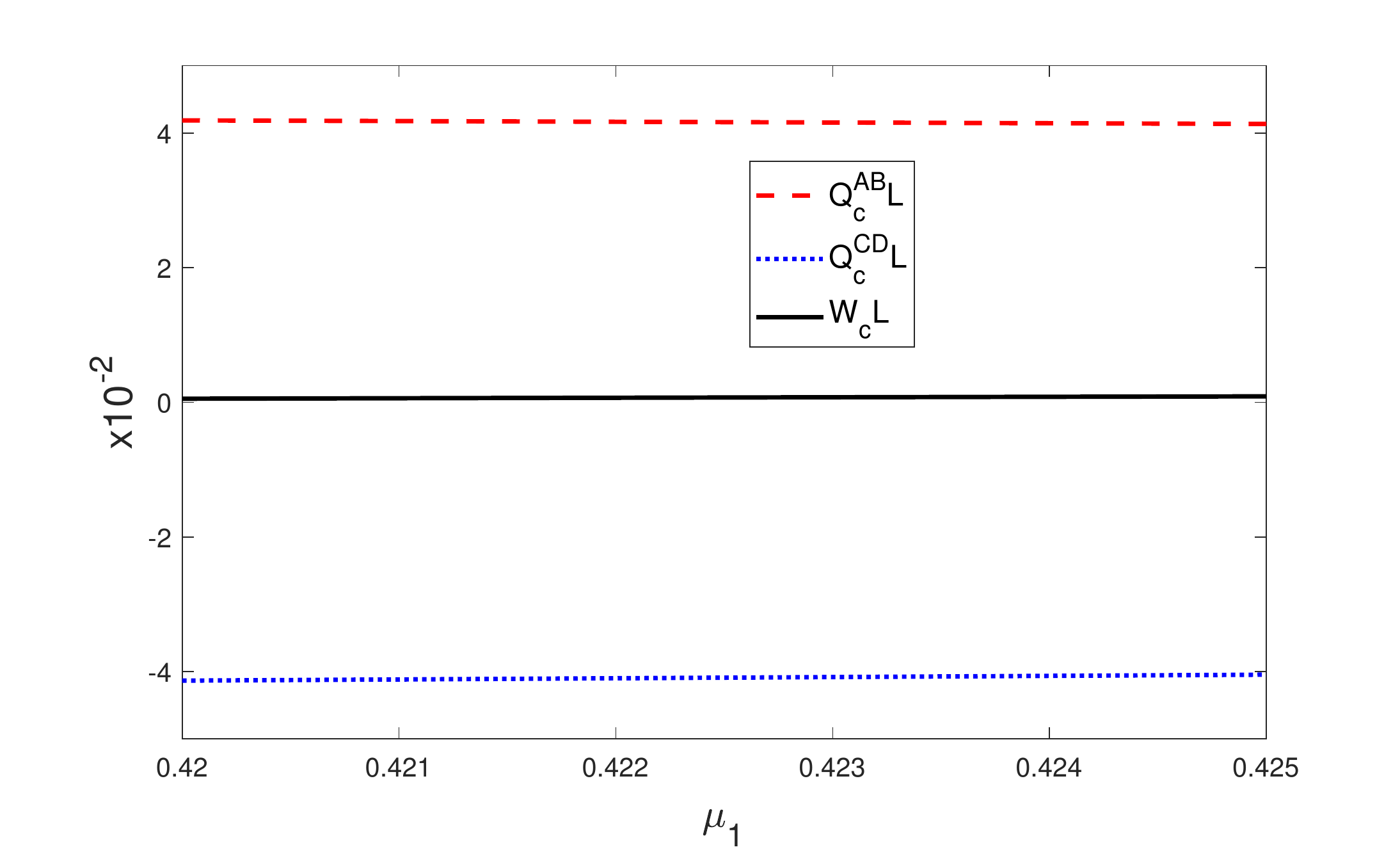}	    \label{fig:6a}}\qquad
	\subfloat[]{\includegraphics[width=8.3cm,height=5.9cm]{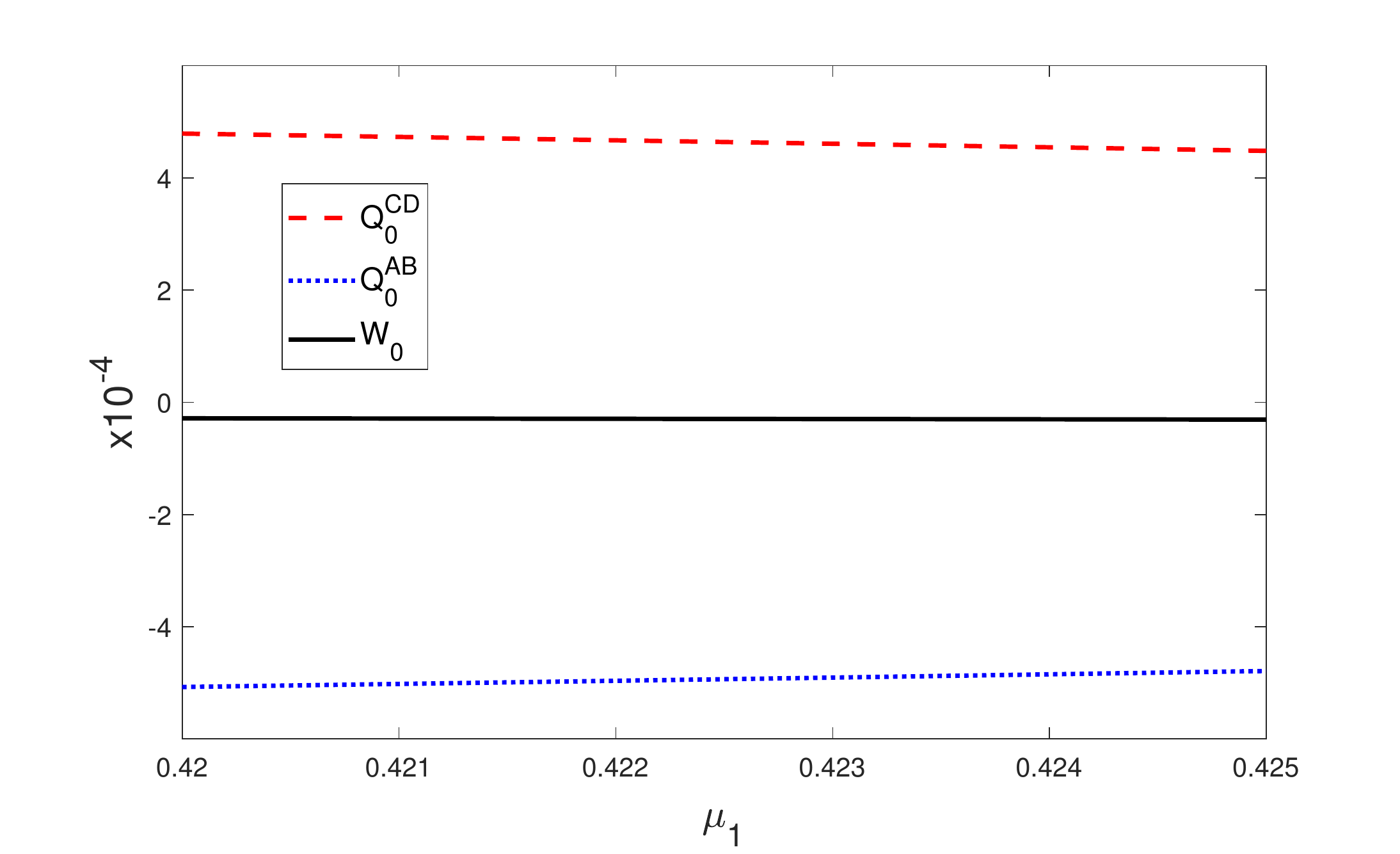}	\label{fig:6b}}
	\caption{(Color online)~ The injected heat $Q_c^{\text{AB}}L$, $Q_0^{\text{CD}}$ (red dashed) and  ejected heat, $Q_c^{\text{CD}}$, $Q_0^{\text{AB}}$ (blue dotted) and  the work output $W_c$, $W_0$ (black solid) corresponding to (a) the bulk heat engine and (b) the boundary refrigerator, respectively, as a function of the chemical potential $\mu_1$ of the first isochore of a finite-length Kitaev chain in independent Otto cycles, operating between a hot bath at temperature $T_B= 0.08$ and a cold bath at temperature $T_D = 0.05$ in the parameter range between $\mu_1=0.42$ and $\mu_2=0.425$. Note that the intermediate temperatures for each engine would be different due to different entropy values leading to different isoentropy conditions.}
	\label{fig:6}
\end{figure*}
\begin{figure}[htb!]
	\centering
	\includegraphics[width=8.3 cm,height=5.9cm]{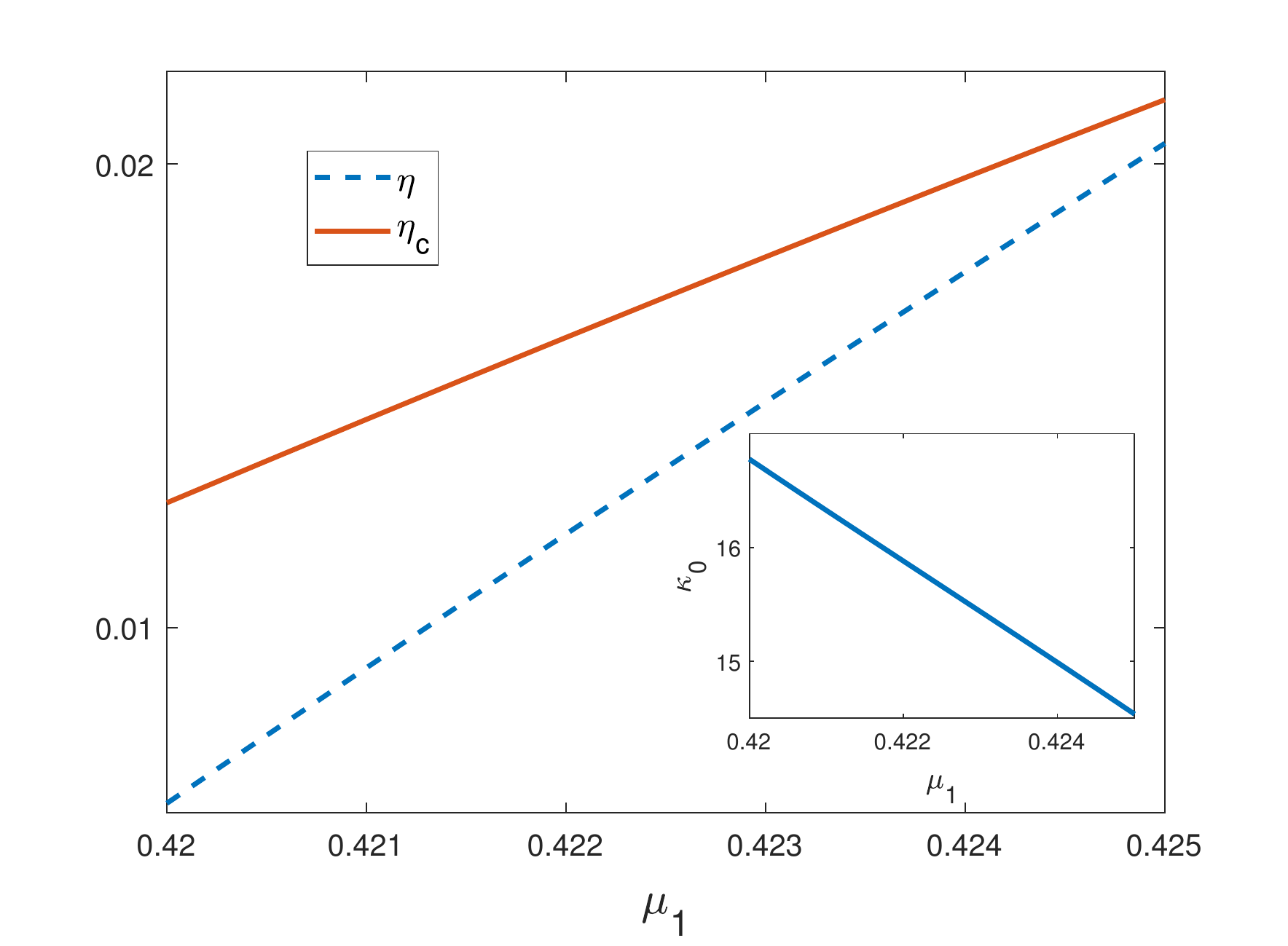}
	\caption{(Color online)~ The efficiencies of the total Otto heat engine $\eta$ (blue dashed) and the bulk Otto heat engine $\eta_c$ (red solid) are given as a function of the chemical potential $\mu_1$ of the cold isochore of a finite-length Kitaev chain in independent Otto cycles, operating between a hot bath at temperature $T_B= 0.08$ and a cold bath at temperature $T_D = 0.08$ in the parameter range between $\mu_1=0.42$ and $\mu_2=0.425$. In the inset, the coefficient of performance of the boundary Otto refrigerator $\kappa_0$ is given.}	
	\label{fig:7}
\end{figure}

\appendix*
\section{Three Otto cycles with a single finite-length Kitaev chain}\label{sec:app}
In this appendix, we investigate whether one can find a parameter range $\mu_1-\mu_2$ and hot and cold bath temperatures, for which bulk, boundary, and total systems could be used as working systems for Otto cycles. When one of them is chosen to work in the Otto cycle however, the other two cannot be found in the Otto cycles.
It is possible to identify three independently running Otto cycles, associated with the total, bulk and boundary of the system separately. For that aim, we examine the curves in the temperature-entropy planes  $T-S$, $T-S_c$, and $T-S_0$ separately. We consider the same hot and cold bath temperatures $T_B=0.08$ and $T_D=0.05$, respectively, for the cycles as in the previous discussions.

The Otto cycles can be determined for the bulk and for the boundary, similarly to the case of the total system. The cycles will be different due to the differences in the intermediate temperatures determined by isentropy conditions for the bulk and boundary entropies $S_c$ and $S_0$, and the entropy of the total chain $S$, and hence would yield different work outputs with different characteristics. In general, our numerical investigations suggest that the intermediate temperatures for the bulk and for the total system are close to each other, while the boundary can attain significantly different intermediate temperatures. We find a narrow regime in the topological phase, between $\mu_1=0.42$ and $\mu_2=0.425$ for which the boundary can produce work through an Otto cycle and act as a refigerator.

The work outputs are shown in Figs.~\ref{fig:6a} and \ref{fig:6b}.  In Fig.~\ref{fig:6a}, we observe that $Q_c^{AB}L>0$ and $Q_c^{CD}L<0$ so that the bulk operation can properly be described as a heat engine with $W_cL>0$. There are heat exchanges only on the isoparametric stages from A to B and C to D, which, in addition makes it an ideal Otto heat engine. We observe in Fig.~\ref{fig:6b} that heat is injected to (ejected from) the boundary of the Kitaev chain, as the system is brought into contact with the cold bath at point D (hot bath at point A) so that $Q_0^{CD}>0 \ (Q_0^{AB}<0)$. This makes the boundary thermal cycle an ideal Otto refrigerator with $W_0<0.$ For the temperature and parameter range considered, the bulk behaves like an Otto heat engine, while the boundary becomes an Otto refigerator.

When we compare efficiencies of the total system and bulk heat engines, $\eta=W/Q_{AB}$ and $\eta_c=W_c/Q_c^{AB}$, respectively, as in Fig.~\ref{fig:7}, we observe that the bulk is more efficient than the total system.  The coefficient of performance of the boundary refrigerator $\kappa_0=Q_0^{CD}/|W_c|$ is given in the inset of Fig.~\ref{fig:7}.

\newpage


%

\end{document}